\documentclass[preprint,superscriptaddress,showpacs,aps,prl]{revtex4-1}

\usepackage{graphicx}
\usepackage{amsmath}
\usepackage{amssymb}
\usepackage{subfigure}
\usepackage{color}
\usepackage{ccaption}
\captionstyle{\raggedright}

\usepackage{ulem}

\renewcommand{\emph}[1]{{\textit{#1}}}

\newcommand{\ie}{\textit{i.e.}}

\newcommand{\dissip}{\varepsilon}
\newcommand{\VA}{\mathbf{V}\cdot\mathbf{a}}

\begin{document} 

\title{Power Fluctuations and Irreversibility in Turbulence}

\author{Haitao Xu}
\email{haitao.xu@ds.mpg.de}
\affiliation{International Collaboration for Turbulence Research}
\affiliation{Max Planck Institute for Dynamics and Self-Organization (MPIDS), G\"ottingen, D-37077 Germany}

\author{Alain Pumir}
\email{alain.pumir@ens-lyon.fr}
\affiliation{International Collaboration for Turbulence Research}
\affiliation{Laboratoire de Physique, Ecole Normale Sup\'erieure de Lyon, Universit\'e Lyon 1 and CNRS, F-69007 France}

\author{Gregory Falkovich}
\affiliation{International Collaboration for Turbulence Research}
\affiliation{Physics of Complex Systems, Weizmann Institute of Science, Rehovot 76100, Israel  and Institute for Information Transmission Problems, Moscow, 127994 Russia}

\author{Eberhard Bodenschatz}
\affiliation{International Collaboration for Turbulence Research}
\affiliation{Max Planck Institute for Dynamics and Self-Organization (MPIDS), G\"ottingen, D-37077 Germany}
\affiliation{Institute for Nonlinear Dynamics, University of G\"ottingen, D-37077 Germany}
\affiliation{Laboratory of Atomic and Solid State Physics and Sibley School of Mechanical and Aerospace Engineering, Cornell University, Ithaca, NY 14853, USA}

\author{Michael Shats}
\affiliation{Research School of Physics and Engineering, The Australian National University, Canberra ACT 0200, Australia}

\author{Hua Xia}
\affiliation{Research School of Physics and Engineering, The Australian National University, Canberra ACT 0200, Australia}

\author{Nicolas Francois}
\affiliation{Research School of Physics and Engineering, The Australian National University, Canberra ACT 0200, Australia}

\author{Guido Boffetta}
\affiliation{International Collaboration for Turbulence Research}
\affiliation{Department of Physics and INFN, University of Turin, I-10125 Torino, Italy}

\date{\today}

\begin{abstract}
The breaking of detailed balance, the symmetry between forward and backward probability transition between two states, is crucial to understand irreversible systems. In hydrodynamic turbulence, a far-from equilibrium system, we observe a strong manifestation of the breaking of detailed balance by following the evolution of the kinetic energy of individual fluid elements. We found in all the flows that we have investigated that fluid elements decelerate faster than they accelerate, giving rise to negative third moment of energy increments, independently of space dimensionality. The exchange of energy between fluid elements however is fundamentally different in two and three dimensions. 
While pressure forces 
do not provide net energy change to slow or fast particles in two dimensions, 
they tend to transfer energy from {\it slow} to {\it fast} particles in three dimensions, possibly implying a runaway of energy.
\end{abstract}


\maketitle

In thermal equilibrium, detailed balance makes equal the probabilities of forward and backward transitions between any states, ensuring time-reversibility of equilibrium statistics \cite{onsager:1931,pine:2005}.
In non-equilibrium, this detailed balance is violated and the statistics is not time-reversible.
Progress in understanding general features of non-equilibrium has been hindered by the lack of a quantitative measure of time irreversibility.
Here we show that for fluid turbulence, a paradigm for ultimate far-from-equilibrium states \cite{Falkovich:2006}, irreversibility manifests itself in the evolution of the kinetic energy of individual fluid elements.
In all flows studied \cite{XPB11,li:2008,xia:2009,francois:2013,BM10} we found that fluid elements decelerate faster than accelerate, a feature known all too well from driving in dense traffic \cite{Helbing:2001}. This asymmetry gives rise to negative third moment of energy changes of fluid  elements. The moment remains constant for time delays in the range characteristic of turbulent eddies, independently of the flow details including space dimensionality.
However, turbulence in two and three dimensions show striking differences in how energy is exchanged between fluid elements: 
pressure forces provide no net energy exchange between fast and slow fluid elements in two dimensions; 
conversely, in three dimensions, pressure transfers energy from {\it slow} to {\it fast} ones.
This suggests the possibility of a runaway of kinetic energy, which may need to be considered when addressing the ``Clay Institute Millennium Problem'' on the three-dimensional Navier-Stokes equations \cite{fefferman:2006}.
In addition, we expect that our approach can be applied to plasma turbulence \cite{krommes:2002}, quantum turbulence \cite{PL11}, magnetohydrodynamics \cite{verma:2004}, and any system that is irreversible and has a separation of scales.
Investigation of such systems in the spirit of the present work is likely to lead to new concepts in the physics of non-equilibrium \cite{derrida:2007}.

Turbulence features originate from the vastly different scales $l_F$, where the flow is forced and inertia dominates, to a scale $l_D$, where dissipation takes over.
For a balance between forcing and dissipation in a statistically steady flow, on average energy is transferred through scales at a rate $\dissip$, a phenomenon called energy cascade.
In three-dimensional (3D) flows, where $l_F \gg l_D$ \cite{frisch:1995,pope:2000,Falkovich:2006}, energy cascades from large to small scales. Energy transfers from small to large scales in two-dimensional (2D) flows, where $l_F \ll l_D$~\cite{K67,BE12}. The energy flux must be a source of statistical irreversibility.
It was suggested that ``A trained eye viewing a movie of steady turbulence run backwards, can tell that something is indeed wrong!'' \cite{Falkovich:2006}.
Yet the rate of energy cascade $\dissip$ is a dimensional quantity so it can be arbitrarily large by changing the units even if the system is very close to equilibrium. Moreover, it can be expressed as a moment of velocity
differences at a single time \cite{frisch:1995,pope:2000}
without any reference to the evolution of the flow. That strongly suggests that $\dissip$ cannot in itself be a proper quantity that characterizes irreversibility.

Here we show that the irreversibility of the flow is
revealed by changing the paradigm from energy redistribution between
``scales'' to  the dynamics of energy changes of fluid elements (or ``particles'').  Such particle (Lagrangian) perspective has recently provided much new insight into the properties of fluid turbulence \cite{laPorta01,yeung:2002,BOX06,XPB11}.
However, characterizing time irreversibility from the motion of a single particle has been impossible so far, as the concepts used to study turbulent
motions were essentially based on the velocity changes along trajectories,
$\mathbf{V}(t) - \mathbf{V}(0)$, whose statistical properties are invariant under the
$t \rightarrow -t$ transformation \cite{FXPB3LT}.
Here we find that the most simple and fundamental scalar quantity, the kinetic energy of a fluid particle, $E(t) = \frac{1}{2}\mathbf{V}^2(t)$, provides a clear identification and quantification of the irreversibility of the flow.
Exemplarily, the reason can be best understood from
Figures~\ref{fig:KE_assym}a \& \ref{fig:KE_assym}b, which show the evolution
of $E(t)$ along the trajectory of a fluid particle in a
3D laboratory water flow \cite{BOX06,XPB11}. It illustrates
that it takes a longer time to build-up than to dissipate a large change of particle kinetic energy. This feature is reminiscent of what has been observed in very
different systems, such as cars in traffic \cite{Helbing:2001,KKH06}, jamming \cite{LN98} and even fluctuations of stock values \cite{JJS03}.
As a result, the statistics of the energy increments $W(\tau)= E(t+\tau)-E(t)$  are
skewed: odd moments are negative for $\tau > 0$.
Figure~\ref{fig:KE_assym}c shows $-W^3(\tau)$ measured from both experiments and
numerical simulations \cite{XPB11,li:2008} of 3D turbulence.
In all these flows, $-W^3(\tau)$ grows at short times and then remains
approximately constant over the entire range of turbulence dynamical time scales.
This contrasts sharply with the lack of scaling range in the
statistics of $\mathbf{V}(t+\tau)- \mathbf{V}(t)$ available so far \cite{yeung:2002,SY11}. 
We note that the kinetic energy $E(t)$ is not Galilean invariant.

We analyzed large experimental and numerical data sets \cite{XPB11,li:2008,xia:2009,francois:2013,BM10} containing more than $10^8$ statistical samples, both in 2D and 3D. Figure~\ref{fig:KE_assym}d shows that the third moment of the energy increments $W(t)$ in 2D is similar to those in 3D (Fig.~\ref{fig:KE_assym}c), \ie, independent of the direction of the energy flux.
This stresses again that the energy flux $\dissip$ by itself is not an appropriate characteristic of irreversibility and
suggests to use instead the dimensionless rate of change of the kinetic energy as a proper measure.
More specifically, we consider here the power normalized by the energy flux,
$p/\dissip$, where $p = \lim\limits_{\Delta t \rightarrow 0}\frac{W(\Delta t)}{\Delta t} = \frac{\mathrm{d}E}{\mathrm{d}t} = \mathbf{V}\cdot \mathbf{a}$, with $\mathbf{a} = \frac{\mathrm{d}\mathbf{V}}{\mathrm{d}t}$ being the fluid acceleration.
We note that the statistics of the power $p$ may be affected by specific, non-universal aspects of the forcing, especially in 2D, in which the external forcing acts at small scales and is fast-changing. As we are focusing on the universal properties of turbulence, in 2D we consider only kinetic energy differences over time scales $\Delta t$ which
are very short in terms of turbulence dynamics, but are
long compared to the characteristic time of the forcing (see more discussion in Appendix).

Time reversibility would imply detailed balance in the sense that the probability of energy gain ($p>0$) is the
same as the probability of energy loss ($p<0$) for any particle with any velocity.
Asymmetric (skewed) PDFs of $p$, as shown in Fig.~\ref{fig:KE_assym}e \& \ref{fig:KE_assym}f, are therefore a signature that detailed balance is violated.
This violation is then quantified by the third moment of the fluctuations of
$p$, which changes sign when reversing $t \rightarrow -t$, thus enabling
a detection of whether the movie of turbulence is played backwards or forwards, and
provides significant new information on the dynamics of turbulence \cite{pomeau:1982}.

Figures~\ref{fig:W_Stat}a \& \ref{fig:W_Stat}b show the normalized third moment of the power, defined as
\begin{equation}
	Ir = -\langle p^3 \rangle/ \dissip^3 ,
\label{eq:Ir}
\end{equation}
in both 2D and 3D.
As described below, $Ir$ is a true measure of irreversibility. It increases with the separation of scales between forcing and dissipation, characterized by the Reynolds numbers.
In 3D, it grows approximately as $Ir \propto R_\lambda^2$, where $R_\lambda \propto (l_F/l_D)^{2/3}$ is the Taylor-scale Reynolds number for 3D turbulence.
For 2D turbulence in the energy cascade regime, \ie, $l_F \ll l_D$, we characterize the scale separation by a friction based Reynolds number $R_\alpha \propto (l_D/l_F)^{2/3} $. 
Figure~\ref{fig:W_Stat}b shows that data from both experiments and numerics suggest that the irreversibility also grows with this Reynolds number approximately as:
$ Ir \propto R_\alpha^2$.
The second moment of $p$, $\langle p^2 \rangle /\dissip^2$, grows with the Reynolds numbers as $R_\alpha^{4/3}$ and $R_\lambda^{4/3}$, as shown in Figures~\ref{fig:W_Stat}c \& \ref{fig:W_Stat}d.
As a consequence, the skewness of the power fluctuations, defined as
$s= \langle p^3 \rangle/ \langle p^2 \rangle^{3/2}$, is approximately constant,
independent of the Reynolds number in both 2D and 3D.
Our results indicate that the PDFs of $p$, however, are not strictly self-similar, and depend on the Reynolds number.

Thus, remarkably, the measure of irreversibility, $Ir$, directly accessible in laboratory flows, depends on
the Reynolds number (or the ratio of the forcing and dissipation scales) to a simple power,
but is independent of the specificity of the forcing.
Even more surprisingly, the dependence of $Ir$ on the Reynolds number is qualitatively very similar in 2D and 3D flows, in spite of the opposite directions of the energy flux.
The profound difference between 2D and 3D turbulence, however, manifests itself in the way the different forces in the fluid act on particles, as revealed by the following analysis. From the Navier-Stokes equations, the power $p$ is expressed as:
\begin{equation}
p=\mathbf{V} \cdot \mathbf{a} = \mathbf{V}
\cdot ( \mathbf{f} - \nabla P + \mathbf{D})
\end{equation}
where $\mathbf{f}$ is the external force per unit mass, $P$ is the pressure divided by density, $\mathbf{D}$ is the
dissipation, due to viscosity in three dimensions,
 ($\mathbf{D} = \nu \nabla^2 \mathbf{V}$), and including also
friction in two dimensions,
($\mathbf{D} = \nu \nabla^2 \mathbf{V} - \alpha \mathbf{V}$).
In a statistically homogeneous, isotropic and steady flow, power is a fluctuating quantity whose mean value
is zero: on average, forcing and dissipation balance each other
$\langle \mathbf{V} \cdot \mathbf{f} \rangle = - \langle \mathbf{V} \cdot \mathbf{D} \rangle = \dissip $.
The pressure force  does no
work on average~\cite{landau_FM}, $\langle \mathbf{V} \cdot \nabla P \rangle = 0$, and merely
redistributes energy between fluid particles.
Both 2D and 3D numerical simulations show that
$ -\mathbf{V} \cdot \nabla P$ provides the main contribution to the variance of $p$, which vastly exceeds the forcing or dissipation terms
in Eq.~(2), see Fig. \ref{fig:NS_contrib}.

Contrary to the variances, the third moments of $p$ in 2D and 3D are dominated by different components of the forces.
In 2D, nearly $60$\% of the power skewness comes from the pressure-work skewness,
$ - \langle ( \mathbf{V} \cdot \nabla P)^3 \rangle$, which is negative.
In other words, the dynamics of fluid elements is dominated by pressure, which essentially
determines the statistics of large fluctuations and the degree of turbulence irreversibility.
The correlation between the pressure gradient term and the friction term 
$\langle 3(-V\cdot \nabla P)^2 )(-\alpha V \cdot V) \rangle$ contributes another $40$\%.
Fig.~\ref{fig:pressure}a shows that the conditional mean
$\langle - \mathbf{V} \cdot \nabla P | \mathbf{V}^2 \rangle$ 
is essentially zero for all values of $\mathbf{V}^2$, which is the most straightforward way to
satisfy the constraint $\langle \mathbf{V} \cdot \nabla P \rangle = 0$.
In 3D, however, it is dramatically different. The numerical simulation at $R_\lambda = 430$ gives the
skewness of power $s = -0.667$, while the pressure contribution
$-\langle (\mathbf{V} \cdot \nabla P)^3 \rangle/ \langle p^2 \rangle^{3/2}=0.023$
is very weak and with a sign opposite to the sign of $s$.
The main contribution to the negative third moments in 3D flows is given by
the cross-correlation of the pressure with dissipation, which dominates the positive contributions from the pressure term and the pressure-forcing correlation term.
The surprising role of pressure in 3D turbulence is clearly
seen from Fig.~\ref{fig:pressure}b, which shows that the conditional mean
$\langle - \mathbf{V} \cdot \nabla P | \mathbf{V}^2 \rangle $ is negative for
small $\mathbf{V}^2$ and positive for large $\mathbf{V}^2$, see also
recent results \cite{wilczek:2011}.
That is, in 3D turbulence, pressure takes energy away from \textit{slow}
particles and gives it to \textit{fast} particles!
By itself, the dynamics due to pressure forces would lead to the runaway of the kinetic energy of some fluid elements, a feature likely related to the accelerated nature of the turbulence energy cascade in three dimensions \cite{Falkovich:2006}.
This runaway provides a physical mechanism that could lead to an unbounded growth of the velocity, a necessary condition for breaking the
regularity of the solution of the Navier-Stokes equation \cite{Leray:1934}.
Therefore our finding could provide important insight on the corresponding Clay Institute Millennium Problem, for which, as explicitly stated in Ref. \cite{fefferman:2006}, ``some deep, new ideas are needed''.

\section*{Ackowledgment}
We thank the Kavli Institutes for Theoretical Physics, where the work started during the 2011 program ``The nature of turbulence" (KITP, Santa Barbara, USA) and continued during the 2012 program ``New directions in turbulence'' (KITPC, Beijing, China). We also acknowledge partial support from Max Planck Society and the EU COST Action MP0806 ``Particles in turbulence''. The work was supported by the grants of German Science Foundation (XU/91-3), Israel Science Foundation and Bi-National Science Foundation in Israel, ANR grant TEC2, the Australian Research Council Discovery Projects funding scheme (DP110101525) and Discovery Early Career Research Award (DE120100364).

H.Xu, A.P., G.F. and E.B. designed research and wrote the manuscript. H.Xu and E.B. designed and performed the three-dimensional experiments and analysed the data. A.P. designed and performed three-dimensional numerical modelling and analysed data with H.Xu.  M.S., H.Xia, and N. F. designed and performed the two-dimensional experiments and analysed data with G.F.. G.B. designed and performed two-dimensional numerical modelling and analysed the data. All authors discussed the physics and commented on the manuscript.

\section*{Appendix: Influence of the forcing on the power statistics}

Turbulence generally depends on the nature of forcing. Of most fundamental interest are the properties that are universal \ie, independent of forcing.

In 3D, a large-scale force impacts strongly on the single-time statistics in
the inertial interval via intermittency and anomalous scaling \cite{Falkovich:2006}. On the
other hand, the moments of the forcing term, $\mathbf{f} \cdot \mathbf{V}$,
behave as: $\langle (\mathbf{f} \mathbf{V})^n \rangle \sim \varepsilon^n$
as a function of the order $n$ of the moment. The
reason for this is that both $\mathbf{V}$ and $\mathbf{f}$ are quantities
that vary on a large scale. In particular, the fluctuations of $\mathbf{f}$
are small compared to the fluctuations of $\mathbf{D}$ and $\nabla P$.
As a consequence, the contribution of
$\langle (\mathbf{f} \cdot \mathbf{V})^3 \rangle$ to the third moment of power
$\langle (\mathbf{V} \cdot \mathbf{a})^3 \rangle$ is small: the statistics
of power are unaffected by the details of the forcing.

In 2D, a small-scale force may affect significantly the
statistics of $\mathbf{V} \cdot \mathbf{a}$. The reason is that due to the
inverse cascade, the forcing $\mathbf{f}$ varies on small scales, in contrary to the velocity field $\mathbf{V}$ that is varying on large
scales. The scalar product $\mathbf{f} \cdot \mathbf{V}$ then rapidly varies along a particle trajectory.
Thus, the averaged quantity $ \langle \mathbf{f} \cdot \mathbf{V} \rangle$
involves many cancellations. In particular, one could conceivably have
$ \dissip = \langle ( \mathbf{f} \cdot \mathbf{V} ) \rangle
\ll \langle ( \mathbf{f} \cdot \mathbf{V} )^2 \rangle^{1/2}$,
and similarly,
$\langle ( \mathbf{f} \cdot \mathbf{V})^3 \rangle \gg \dissip^3$,
implying that the details of the forcing may be important while studying
the third moment of $\VA$.

In the results from 2D numerical simulations in which white-noise forcing is implemented, the sign of the third moment of $\VA$ is negative.
On the other hand, in turbulent flows induced by surface-waves, the forcing is well correlated in time due to the coherency of the monochromatic Faraday waves at the frequency of $f_F=f_0/2$. This leads to a positive sign of $\langle (\mathbf{V}\cdot \mathbf{a})^3 \rangle$. Due to this nature of turbulence forcing in these experiments, fluid particles are accelerated fast by the waves at the time scale of $T_F=f_F^{-1}$, and then slowly lose energy via turbulence interactions. This property is a manifestation of the non-universal nature of forcing in 2D. However, since the time scales over which turbulent dynamics develop are much larger than $T_F$, averaging $\VA$ in time along the trajectories over several periods of Faraday waves, \ie, defining $p(t) = \frac{1}{sm} \int_0^{sm} \mathbf{V}(t+t') \cdot \mathbf{a}(t+t') dt'$, where $sm$ is the time period of averaging, eliminates the contribution of forcing on the $\VA$ statistics and reveals negative $\langle p^3 \rangle$ in 
 the inertial interval. Similarly, we filter the energy increments along the trajecotries as $W(\tau) = \frac{1}{sm} \int_0^{sm} \frac{1}{2}[\mathbf{V^2}(t+\tau+t') - \mathbf{V}^2(t+t')] dt'$. On the other hand, the peaks at $\approx f_F$ are much more prominent in the spectra of $\mathbf{V} \cdot \mathbf{a}$.

The contribution of forcing can be seen in the Lagrangian spectra of $\mathbf{V^2}$ and  $\mathbf{V} \cdot \mathbf{a}$ computed along the tracer trajectories in the surface wave driven 2D turbulence. As shown in Figs.~\ref{fig:2d_turb_av_smooth}a \& \ref{fig:2d_turb_av_smooth}b, the spectra exhibit strong peaks at a frequency close to $f_F $, corresponding to the Doppler shifted Faraday wave frequency. Another peak seen at lower frequencies ($f \leq 6$ Hz) corresponds to the turbulent motion, including the inertial range.
Low-pass filtering $\mathbf{V^2}$, more precisely, averaging $\mathbf{V^2}$ over $sm = 2 T_F$ along the trajectories as defined above, eliminates the forcing peak in the spectra of $\mathbf{V^2}$, see Fig.~\ref{fig:2d_turb_av_smooth}a.
The data shown in Fig.~1d of the main text are obtained after applying this filtering with $sm = 2 T_F$. On the other hand, the peaks at $\approx f_F$ are much more prominent in the spectra of $\mathbf{V} \cdot \mathbf{a}$ and contribute to a strong signal, which is responsible for the measured positive skewness of $\langle (\mathbf{V}\cdot\mathbf{a})^3 \rangle$.
Although smoothing $\mathbf{V}\cdot\mathbf{a}$ along the Lagrangian trajectories over $2 T_F$ reduces significantly the relative contribution of the peak at $\approx f_F$, see Fig.~\ref{fig:2d_turb_av_smooth}b, the filtered signal still exhibits a positive skewness.
In Fig.~\ref{fig:2d_turb_av_smooth}c, we show the skewness of the averaged power $p$ as a function of $sm$, the time of averaging. As the filtering period is increased, the third moment $\langle p^3 \rangle$ decreases and reaches an essentially constant negative value when the smoothing time is larger than $\sim 5 T_F$, see Fig.~\ref{fig:2d_turb_av_smooth}c. A similar behavior is observed for all values of the vertical accelerations.
We note that because $W$ and $p$ are obtained by smoothing $\mathbf{V}^2$ and $\VA$ separately, the relation $p = \frac{\Delta W}{\Delta t}$ is only approximately held due to the limited sampling rate of the cameras. On the other hand, this does not affect our results since we are considering the turbulent dynamics, which, as shown in Figs.~\ref{fig:2d_turb_av_smooth}a \& \ref{fig:2d_turb_av_smooth}b, are in the smaller frequency range. Over that range, the skewness of the smoothed power $\langle p^3 \rangle$ is independent of the averaging time $sm$.
In Figs.~2b \& 2d in the main text the second and the third moments of $p$ from experimental data are computed by smoothing $p$ over $7 T_F$.

\bibliography{ir_ref}

\begin{thebibliography}{30}%
\makeatletter
\providecommand \@ifxundefined [1]{%
 \@ifx{#1\undefined}
}%
\providecommand \@ifnum [1]{%
 \ifnum #1\expandafter \@firstoftwo
 \else \expandafter \@secondoftwo
 \fi
}%
\providecommand \@ifx [1]{%
 \ifx #1\expandafter \@firstoftwo
 \else \expandafter \@secondoftwo
 \fi
}%
\providecommand \natexlab [1]{#1}%
\providecommand \enquote  [1]{``#1''}%
\providecommand \bibnamefont  [1]{#1}%
\providecommand \bibfnamefont [1]{#1}%
\providecommand \citenamefont [1]{#1}%
\providecommand \href@noop [0]{\@secondoftwo}%
\providecommand \href [0]{\begingroup \@sanitize@url \@href}%
\providecommand \@href[1]{\@@startlink{#1}\@@href}%
\providecommand \@@href[1]{\endgroup#1\@@endlink}%
\providecommand \@sanitize@url [0]{\catcode `\\12\catcode `\$12\catcode
  `\&12\catcode `\#12\catcode `\^12\catcode `\_12\catcode `\%12\relax}%
\providecommand \@@startlink[1]{}%
\providecommand \@@endlink[0]{}%
\providecommand \url  [0]{\begingroup\@sanitize@url \@url }%
\providecommand \@url [1]{\endgroup\@href {#1}{\urlprefix }}%
\providecommand \urlprefix  [0]{URL }%
\providecommand \Eprint [0]{\href }%
\providecommand \doibase [0]{http://dx.doi.org/}%
\providecommand \selectlanguage [0]{\@gobble}%
\providecommand \bibinfo  [0]{\@secondoftwo}%
\providecommand \bibfield  [0]{\@secondoftwo}%
\providecommand \translation [1]{[#1]}%
\providecommand \BibitemOpen [0]{}%
\providecommand \bibitemStop [0]{}%
\providecommand \bibitemNoStop [0]{.\EOS\space}%
\providecommand \EOS [0]{\spacefactor3000\relax}%
\providecommand \BibitemShut  [1]{\csname bibitem#1\endcsname}%
\let\auto@bib@innerbib\@empty
\bibitem [{\citenamefont {Onsager}(1931)}]{onsager:1931}%
  \BibitemOpen
  \bibfield  {author} {\bibinfo {author} {\bibfnamefont {L.}~\bibnamefont
  {Onsager}},\ }\href@noop {} {\bibfield  {journal} {\bibinfo  {journal} {Phys.
  Rev.}\ }\textbf {\bibinfo {volume} {37}},\ \bibinfo {pages} {405} (\bibinfo
  {year} {1931})}\BibitemShut {NoStop}%
\bibitem [{\citenamefont {Pine}\ \emph {et~al.}(2005)\citenamefont {Pine},
  \citenamefont {Gollub}, \citenamefont {Brady},\ and\ \citenamefont
  {Leshansky}}]{pine:2005}%
  \BibitemOpen
  \bibfield  {author} {\bibinfo {author} {\bibfnamefont {D.~J.}\ \bibnamefont
  {Pine}}, \bibinfo {author} {\bibfnamefont {J.~P.}\ \bibnamefont {Gollub}},
  \bibinfo {author} {\bibfnamefont {J.~F.}\ \bibnamefont {Brady}}, \ and\
  \bibinfo {author} {\bibfnamefont {A.~M.}\ \bibnamefont {Leshansky}},\
  }\href@noop {} {\bibfield  {journal} {\bibinfo  {journal} {Nature}\ }\textbf
  {\bibinfo {volume} {438}},\ \bibinfo {pages} {997} (\bibinfo {year}
  {2005})}\BibitemShut {NoStop}%
\bibitem [{\citenamefont {Falkovich}\ and\ \citenamefont
  {Sreenivasan}(2006)}]{Falkovich:2006}%
  \BibitemOpen
  \bibfield  {author} {\bibinfo {author} {\bibfnamefont {G.}~\bibnamefont
  {Falkovich}}\ and\ \bibinfo {author} {\bibfnamefont {K.}~\bibnamefont
  {Sreenivasan}},\ }\href@noop {} {\bibfield  {journal} {\bibinfo  {journal}
  {Phys. Today}\ }\textbf {\bibinfo {volume} {59}},\ \bibinfo {pages} {43}
  (\bibinfo {year} {2006})}\BibitemShut {NoStop}%
\bibitem [{\citenamefont {Xu}\ \emph {et~al.}(2011)\citenamefont {Xu},
  \citenamefont {Pumir},\ and\ \citenamefont {Bodenschatz}}]{XPB11}%
  \BibitemOpen
  \bibfield  {author} {\bibinfo {author} {\bibfnamefont {H.}~\bibnamefont
  {Xu}}, \bibinfo {author} {\bibfnamefont {A.}~\bibnamefont {Pumir}}, \ and\
  \bibinfo {author} {\bibfnamefont {E.}~\bibnamefont {Bodenschatz}},\
  }\href@noop {} {\bibfield  {journal} {\bibinfo  {journal} {Nature Phys.}\
  }\textbf {\bibinfo {volume} {7}},\ \bibinfo {pages} {709} (\bibinfo {year}
  {2011})}\BibitemShut {NoStop}%
\bibitem [{\citenamefont {Li}\ \emph {et~al.}(2008)\citenamefont {Li},
  \citenamefont {Perlman}, \citenamefont {Wan}, \citenamefont {Yang},
  \citenamefont {Meneveau}, \citenamefont {Burns}, \citenamefont {Chen},
  \citenamefont {Szalay},\ and\ \citenamefont {Eyink}}]{li:2008}%
  \BibitemOpen
  \bibfield  {author} {\bibinfo {author} {\bibfnamefont {Y.}~\bibnamefont
  {Li}}, \bibinfo {author} {\bibfnamefont {E.}~\bibnamefont {Perlman}},
  \bibinfo {author} {\bibfnamefont {M.}~\bibnamefont {Wan}}, \bibinfo {author}
  {\bibfnamefont {Y.}~\bibnamefont {Yang}}, \bibinfo {author} {\bibfnamefont
  {C.}~\bibnamefont {Meneveau}}, \bibinfo {author} {\bibfnamefont
  {R.}~\bibnamefont {Burns}}, \bibinfo {author} {\bibfnamefont
  {S.}~\bibnamefont {Chen}}, \bibinfo {author} {\bibfnamefont {A.}~\bibnamefont
  {Szalay}}, \ and\ \bibinfo {author} {\bibfnamefont {G.~L.}\ \bibnamefont
  {Eyink}},\ }\href@noop {} {\bibfield  {journal} {\bibinfo  {journal} {J.
  Turbul.}\ }\textbf {\bibinfo {volume} {9}},\ \bibinfo {pages} {31} (\bibinfo
  {year} {2008})}\BibitemShut {NoStop}%
\bibitem [{\citenamefont {Xia}\ \emph {et~al.}(2009)\citenamefont {Xia},
  \citenamefont {Shats},\ and\ \citenamefont {Falkovich}}]{xia:2009}%
  \BibitemOpen
  \bibfield  {author} {\bibinfo {author} {\bibfnamefont {H.}~\bibnamefont
  {Xia}}, \bibinfo {author} {\bibfnamefont {M.}~\bibnamefont {Shats}}, \ and\
  \bibinfo {author} {\bibfnamefont {G.}~\bibnamefont {Falkovich}},\ }\href@noop
  {} {\bibfield  {journal} {\bibinfo  {journal} {Phys. Fluids}\ }\textbf
  {\bibinfo {volume} {21}},\ \bibinfo {pages} {125101} (\bibinfo {year}
  {2009})}\BibitemShut {NoStop}%
\bibitem [{\citenamefont {Francois}\ \emph {et~al.}(2013)\citenamefont
  {Francois}, \citenamefont {Xia}, \citenamefont {Punzmann},\ and\
  \citenamefont {Shats}}]{francois:2013}%
  \BibitemOpen
  \bibfield  {author} {\bibinfo {author} {\bibfnamefont {N.}~\bibnamefont
  {Francois}}, \bibinfo {author} {\bibfnamefont {H.}~\bibnamefont {Xia}},
  \bibinfo {author} {\bibfnamefont {H.}~\bibnamefont {Punzmann}}, \ and\
  \bibinfo {author} {\bibfnamefont {M.}~\bibnamefont {Shats}},\ }\href@noop {}
  {\bibfield  {journal} {\bibinfo  {journal} {Phys. Rev. Lett.}\ }\textbf
  {\bibinfo {volume} {110}},\ \bibinfo {pages} {194501} (\bibinfo {year}
  {2013})}\BibitemShut {NoStop}%
\bibitem [{\citenamefont {Boffetta}\ and\ \citenamefont
  {Musacchio}(2010)}]{BM10}%
  \BibitemOpen
  \bibfield  {author} {\bibinfo {author} {\bibfnamefont {G.}~\bibnamefont
  {Boffetta}}\ and\ \bibinfo {author} {\bibfnamefont {S.}~\bibnamefont
  {Musacchio}},\ }\href@noop {} {\bibfield  {journal} {\bibinfo  {journal}
  {Phys. Rev. E}\ }\textbf {\bibinfo {volume} {82}},\ \bibinfo {pages} {016307}
  (\bibinfo {year} {2010})}\BibitemShut {NoStop}%
\bibitem [{\citenamefont {Helbing}(2001)}]{Helbing:2001}%
  \BibitemOpen
  \bibfield  {author} {\bibinfo {author} {\bibfnamefont {D.}~\bibnamefont
  {Helbing}},\ }\href@noop {} {\bibfield  {journal} {\bibinfo  {journal} {Rev.
  Mod. Phys.}\ }\textbf {\bibinfo {volume} {73}},\ \bibinfo {pages} {1067}
  (\bibinfo {year} {2001})}\BibitemShut {NoStop}%
\bibitem [{\citenamefont {Fefferman}(2006)}]{fefferman:2006}%
  \BibitemOpen
  \bibfield  {author} {\bibinfo {author} {\bibfnamefont {C.~L.}\ \bibnamefont
  {Fefferman}},\ }\enquote {\bibinfo {title} {The millennium prize problems},}\
  \ (\bibinfo  {publisher} {Clay Mathematics Institute},\ \bibinfo {address}
  {Cambridge, MA},\ \bibinfo {year} {2006})\ Chap.\ \bibinfo {chapter}
  {Existence and smoothness of the {Navier-Stokes} equation}, pp.\ \bibinfo
  {pages} {57--67}\BibitemShut {NoStop}%
\bibitem [{\citenamefont {Krommes}(2002)}]{krommes:2002}%
  \BibitemOpen
  \bibfield  {author} {\bibinfo {author} {\bibfnamefont {J.~A.}\ \bibnamefont
  {Krommes}},\ }\href@noop {} {\bibfield  {journal} {\bibinfo  {journal} {Phys.
  Rep.}\ }\textbf {\bibinfo {volume} {260}},\ \bibinfo {pages} {1} (\bibinfo
  {year} {2002})}\BibitemShut {NoStop}%
\bibitem [{\citenamefont {Paoletti}\ and\ \citenamefont
  {Lathrop}(2011)}]{PL11}%
  \BibitemOpen
  \bibfield  {author} {\bibinfo {author} {\bibfnamefont {M.~S.}\ \bibnamefont
  {Paoletti}}\ and\ \bibinfo {author} {\bibfnamefont {D.~P.}\ \bibnamefont
  {Lathrop}},\ }\href@noop {} {\bibfield  {journal} {\bibinfo  {journal} {Annu.
  Rev. Condens. Matter Phys.}\ }\textbf {\bibinfo {volume} {2}},\ \bibinfo
  {pages} {213} (\bibinfo {year} {2011})}\BibitemShut {NoStop}%
\bibitem [{\citenamefont {Verma}(2004)}]{verma:2004}%
  \BibitemOpen
  \bibfield  {author} {\bibinfo {author} {\bibfnamefont {M.~K.}\ \bibnamefont
  {Verma}},\ }\href@noop {} {\bibfield  {journal} {\bibinfo  {journal} {Phys.
  Rep.}\ }\textbf {\bibinfo {volume} {401}},\ \bibinfo {pages} {229} (\bibinfo
  {year} {2004})}\BibitemShut {NoStop}%
\bibitem [{\citenamefont {Derrida}(2007)}]{derrida:2007}%
  \BibitemOpen
  \bibfield  {author} {\bibinfo {author} {\bibfnamefont {B.}~\bibnamefont
  {Derrida}},\ }\href@noop {} {\bibfield  {journal} {\bibinfo  {journal} {J.
  Stat. Mech. Theor. Exp.}\ ,\ \bibinfo {pages} {P07023}} (\bibinfo {year}
  {2007})}\BibitemShut {NoStop}%
\bibitem [{\citenamefont {Frisch}(1995)}]{frisch:1995}%
  \BibitemOpen
  \bibfield  {author} {\bibinfo {author} {\bibfnamefont {U.}~\bibnamefont
  {Frisch}},\ }\href@noop {} {\emph {\bibinfo {title} {Turbulence: The Legacy
  of A.~N.~Kolmogorov}}}\ (\bibinfo  {publisher} {Cambridge University Press},\
  \bibinfo {address} {Cambridge, England},\ \bibinfo {year} {1995})\BibitemShut
  {NoStop}%
\bibitem [{\citenamefont {Pope}(2000)}]{pope:2000}%
  \BibitemOpen
  \bibfield  {author} {\bibinfo {author} {\bibfnamefont {S.~B.}\ \bibnamefont
  {Pope}},\ }\href@noop {} {\emph {\bibinfo {title} {Turbulent Flows}}}\
  (\bibinfo  {publisher} {Cambridge University Press},\ \bibinfo {address}
  {Cambridge, England},\ \bibinfo {year} {2000})\BibitemShut {NoStop}%
\bibitem [{\citenamefont {Kraichnan}(1967)}]{K67}%
  \BibitemOpen
  \bibfield  {author} {\bibinfo {author} {\bibfnamefont {R.~H.}\ \bibnamefont
  {Kraichnan}},\ }\href@noop {} {\bibfield  {journal} {\bibinfo  {journal}
  {Phys. Fluids}\ }\textbf {\bibinfo {volume} {10}},\ \bibinfo {pages} {1417}
  (\bibinfo {year} {1967})}\BibitemShut {NoStop}%
\bibitem [{\citenamefont {Boffetta}\ and\ \citenamefont {Ecke}(2012)}]{BE12}%
  \BibitemOpen
  \bibfield  {author} {\bibinfo {author} {\bibfnamefont {G.}~\bibnamefont
  {Boffetta}}\ and\ \bibinfo {author} {\bibfnamefont {R.~E.}\ \bibnamefont
  {Ecke}},\ }\href@noop {} {\bibfield  {journal} {\bibinfo  {journal} {Annu.
  Rev. Fluid Mech.}\ }\textbf {\bibinfo {volume} {44}},\ \bibinfo {pages} {417}
  (\bibinfo {year} {2012})}\BibitemShut {NoStop}%
\bibitem [{\citenamefont {La~Porta}\ \emph {et~al.}(2001)\citenamefont
  {La~Porta}, \citenamefont {Voth}, \citenamefont {Crawford}, \citenamefont
  {Alexander},\ and\ \citenamefont {Bodenschatz}}]{laPorta01}%
  \BibitemOpen
  \bibfield  {author} {\bibinfo {author} {\bibfnamefont {A.}~\bibnamefont
  {La~Porta}}, \bibinfo {author} {\bibfnamefont {G.~A.}\ \bibnamefont {Voth}},
  \bibinfo {author} {\bibfnamefont {A.~M.}\ \bibnamefont {Crawford}}, \bibinfo
  {author} {\bibfnamefont {J.}~\bibnamefont {Alexander}}, \ and\ \bibinfo
  {author} {\bibfnamefont {E.}~\bibnamefont {Bodenschatz}},\ }\href@noop {}
  {\bibfield  {journal} {\bibinfo  {journal} {Nature}\ }\textbf {\bibinfo
  {volume} {409}},\ \bibinfo {pages} {1017} (\bibinfo {year}
  {2001})}\BibitemShut {NoStop}%
\bibitem [{\citenamefont {Yeung}(2002)}]{yeung:2002}%
  \BibitemOpen
  \bibfield  {author} {\bibinfo {author} {\bibfnamefont {P.~K.}\ \bibnamefont
  {Yeung}},\ }\href@noop {} {\bibfield  {journal} {\bibinfo  {journal} {Annu.
  Rev. Fluid Mech.}\ }\textbf {\bibinfo {volume} {34}},\ \bibinfo {pages} {115}
  (\bibinfo {year} {2002})}\BibitemShut {NoStop}%
\bibitem [{\citenamefont {Bourgoin}\ \emph {et~al.}(2006)\citenamefont
  {Bourgoin}, \citenamefont {Ouellette}, \citenamefont {Xu}, \citenamefont
  {Berg},\ and\ \citenamefont {Bodenschatz}}]{BOX06}%
  \BibitemOpen
  \bibfield  {author} {\bibinfo {author} {\bibfnamefont {M.}~\bibnamefont
  {Bourgoin}}, \bibinfo {author} {\bibfnamefont {N.~T.}\ \bibnamefont
  {Ouellette}}, \bibinfo {author} {\bibfnamefont {H.}~\bibnamefont {Xu}},
  \bibinfo {author} {\bibfnamefont {J.}~\bibnamefont {Berg}}, \ and\ \bibinfo
  {author} {\bibfnamefont {E.}~\bibnamefont {Bodenschatz}},\ }\href@noop {}
  {\bibfield  {journal} {\bibinfo  {journal} {Science}\ }\textbf {\bibinfo
  {volume} {311}},\ \bibinfo {pages} {835} (\bibinfo {year}
  {2006})}\BibitemShut {NoStop}%
\bibitem [{\citenamefont {Falkovich}\ \emph {et~al.}(2012)\citenamefont
  {Falkovich}, \citenamefont {Xu}, \citenamefont {Pumir}, \citenamefont
  {Bodenschatz}, \citenamefont {Biferale}, \citenamefont {Boffetta},
  \citenamefont {Lanotte},\ and\ \citenamefont {Toschi}}]{FXPB3LT}%
  \BibitemOpen
  \bibfield  {author} {\bibinfo {author} {\bibfnamefont {G.}~\bibnamefont
  {Falkovich}}, \bibinfo {author} {\bibfnamefont {H.}~\bibnamefont {Xu}},
  \bibinfo {author} {\bibfnamefont {A.}~\bibnamefont {Pumir}}, \bibinfo
  {author} {\bibfnamefont {E.}~\bibnamefont {Bodenschatz}}, \bibinfo {author}
  {\bibfnamefont {L.}~\bibnamefont {Biferale}}, \bibinfo {author}
  {\bibfnamefont {G.}~\bibnamefont {Boffetta}}, \bibinfo {author}
  {\bibfnamefont {A.~S.}\ \bibnamefont {Lanotte}}, \ and\ \bibinfo {author}
  {\bibfnamefont {F.}~\bibnamefont {Toschi}},\ }\href@noop {} {\bibfield
  {journal} {\bibinfo  {journal} {Phys. Fluids}\ }\textbf {\bibinfo {volume}
  {24}},\ \bibinfo {pages} {055102} (\bibinfo {year} {2012})}\BibitemShut
  {NoStop}%
\bibitem [{\citenamefont {Kerner}\ \emph {et~al.}(2006)\citenamefont {Kerner},
  \citenamefont {Klenov},\ and\ \citenamefont {Hiller}}]{KKH06}%
  \BibitemOpen
  \bibfield  {author} {\bibinfo {author} {\bibfnamefont {B.~S.}\ \bibnamefont
  {Kerner}}, \bibinfo {author} {\bibfnamefont {S.~L.}\ \bibnamefont {Klenov}},
  \ and\ \bibinfo {author} {\bibfnamefont {A.}~\bibnamefont {Hiller}},\
  }\href@noop {} {\bibfield  {journal} {\bibinfo  {journal} {J. Phys. A: Math.
  Gen.}\ }\textbf {\bibinfo {volume} {39}},\ \bibinfo {pages} {2001} (\bibinfo
  {year} {2006})}\BibitemShut {NoStop}%
\bibitem [{\citenamefont {Liu}\ and\ \citenamefont {Nagel}(1998)}]{LN98}%
  \BibitemOpen
  \bibfield  {author} {\bibinfo {author} {\bibfnamefont {A.~J.}\ \bibnamefont
  {Liu}}\ and\ \bibinfo {author} {\bibfnamefont {S.~R.}\ \bibnamefont
  {Nagel}},\ }\href@noop {} {\bibfield  {journal} {\bibinfo  {journal}
  {Nature}\ }\textbf {\bibinfo {volume} {396}},\ \bibinfo {pages} {21}
  (\bibinfo {year} {1998})}\BibitemShut {NoStop}%
\bibitem [{\citenamefont {Jensen}\ \emph {et~al.}(2003)\citenamefont {Jensen},
  \citenamefont {Johansen},\ and\ \citenamefont {Simonsen}}]{JJS03}%
  \BibitemOpen
  \bibfield  {author} {\bibinfo {author} {\bibfnamefont {M.~H.}\ \bibnamefont
  {Jensen}}, \bibinfo {author} {\bibfnamefont {A.}~\bibnamefont {Johansen}}, \
  and\ \bibinfo {author} {\bibfnamefont {I.}~\bibnamefont {Simonsen}},\
  }\href@noop {} {\bibfield  {journal} {\bibinfo  {journal} {Physica A}\
  }\textbf {\bibinfo {volume} {324}},\ \bibinfo {pages} {338} (\bibinfo {year}
  {2003})}\BibitemShut {NoStop}%
\bibitem [{\citenamefont {Sawford}\ and\ \citenamefont {Yeung}(2011)}]{SY11}%
  \BibitemOpen
  \bibfield  {author} {\bibinfo {author} {\bibfnamefont {B.~L.}\ \bibnamefont
  {Sawford}}\ and\ \bibinfo {author} {\bibfnamefont {P.~K.}\ \bibnamefont
  {Yeung}},\ }\href@noop {} {\bibfield  {journal} {\bibinfo  {journal} {Phys.
  Fluids}\ }\textbf {\bibinfo {volume} {23}},\ \bibinfo {pages} {091704}
  (\bibinfo {year} {2011})}\BibitemShut {NoStop}%
\bibitem [{\citenamefont {Pomeau}(1982)}]{pomeau:1982}%
  \BibitemOpen
  \bibfield  {author} {\bibinfo {author} {\bibfnamefont {Y.}~\bibnamefont
  {Pomeau}},\ }\href@noop {} {\bibfield  {journal} {\bibinfo  {journal} {J.
  Phys. France}\ }\textbf {\bibinfo {volume} {43}},\ \bibinfo {pages} {859}
  (\bibinfo {year} {1982})}\BibitemShut {NoStop}%
\bibitem [{\citenamefont {Landau}\ and\ \citenamefont
  {Lifshitz}(1987)}]{landau_FM}%
  \BibitemOpen
  \bibfield  {author} {\bibinfo {author} {\bibfnamefont {L.~D.}\ \bibnamefont
  {Landau}}\ and\ \bibinfo {author} {\bibfnamefont {E.~M.}\ \bibnamefont
  {Lifshitz}},\ }\href@noop {} {\emph {\bibinfo {title} {Fluid Mechanics}}}\
  (\bibinfo  {publisher} {Butterworth-Heinemann},\ \bibinfo {year}
  {1987})\BibitemShut {NoStop}%
\bibitem [{\citenamefont {Wilczek}\ \emph {et~al.}(2011)\citenamefont
  {Wilczek}, \citenamefont {Daitche},\ and\ \citenamefont
  {Friedrich}}]{wilczek:2011}%
  \BibitemOpen
  \bibfield  {author} {\bibinfo {author} {\bibfnamefont {M.}~\bibnamefont
  {Wilczek}}, \bibinfo {author} {\bibfnamefont {A.}~\bibnamefont {Daitche}}, \
  and\ \bibinfo {author} {\bibfnamefont {R.}~\bibnamefont {Friedrich}},\
  }\href@noop {} {\bibfield  {journal} {\bibinfo  {journal} {J. Fluid Mech.}\
  }\textbf {\bibinfo {volume} {676}},\ \bibinfo {pages} {191} (\bibinfo {year}
  {2011})}\BibitemShut {NoStop}%
\bibitem [{\citenamefont {Leray}(1934)}]{Leray:1934}%
  \BibitemOpen
  \bibfield  {author} {\bibinfo {author} {\bibfnamefont {J.}~\bibnamefont
  {Leray}},\ }\href@noop {} {\bibfield  {journal} {\bibinfo  {journal} {Acta
  Mathematica}\ }\textbf {\bibinfo {volume} {63}},\ \bibinfo {pages} {193}
  (\bibinfo {year} {1934})}\BibitemShut {NoStop}%
\end{thebibliography}%

\clearpage

\begin{figure}[h!]
\begin{center}
\subfigure[]{
	\includegraphics[width=0.47\textwidth]{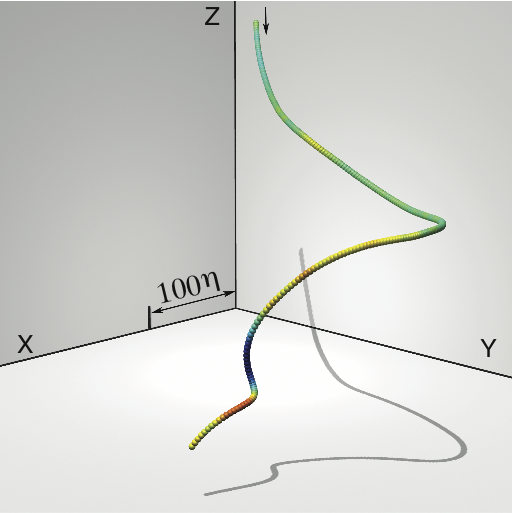}
}
\subfigure[]{
	\includegraphics[width=0.47\textwidth]{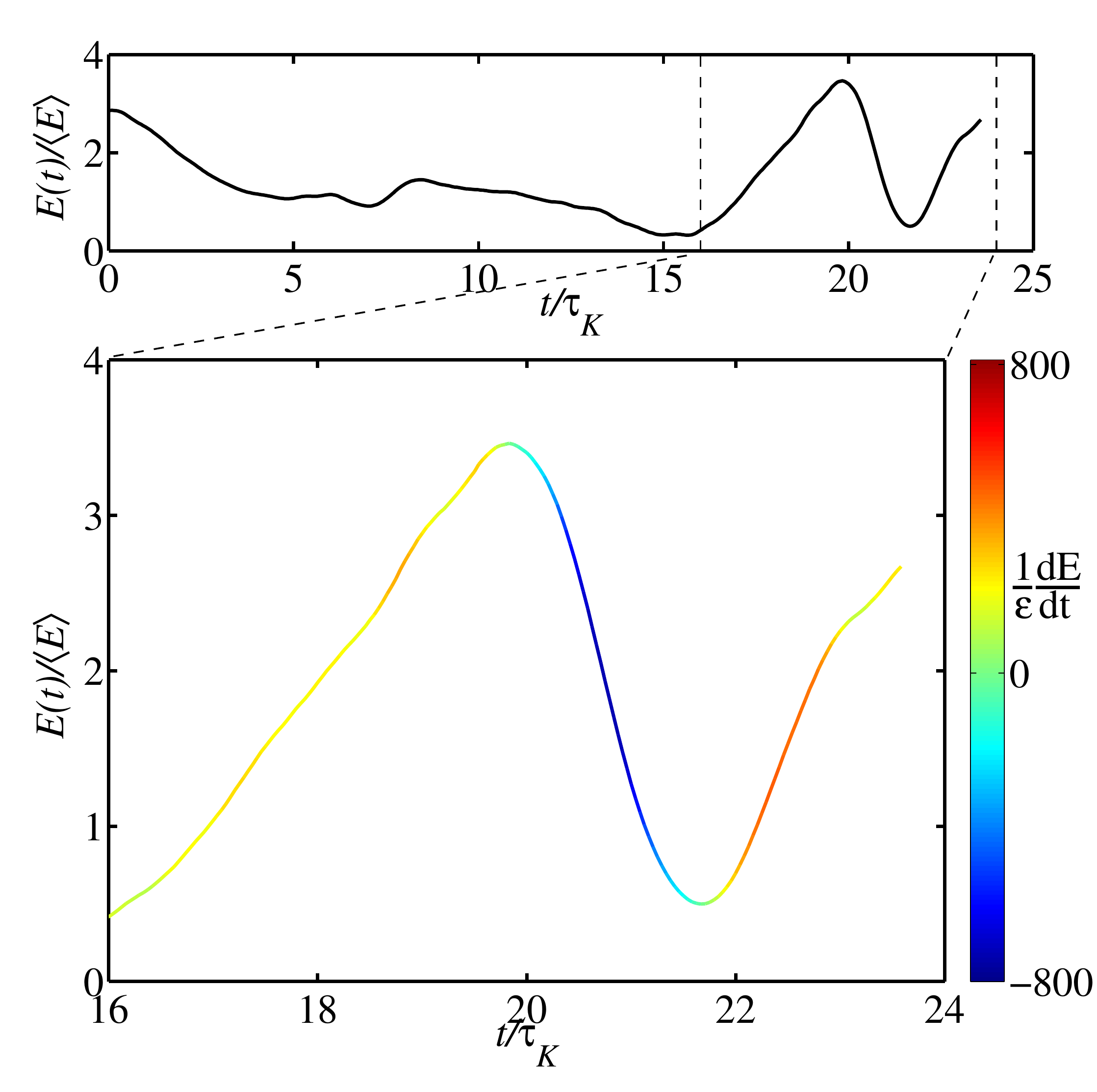}
}
\subfigure[]{
	\includegraphics[width=0.44\textwidth]{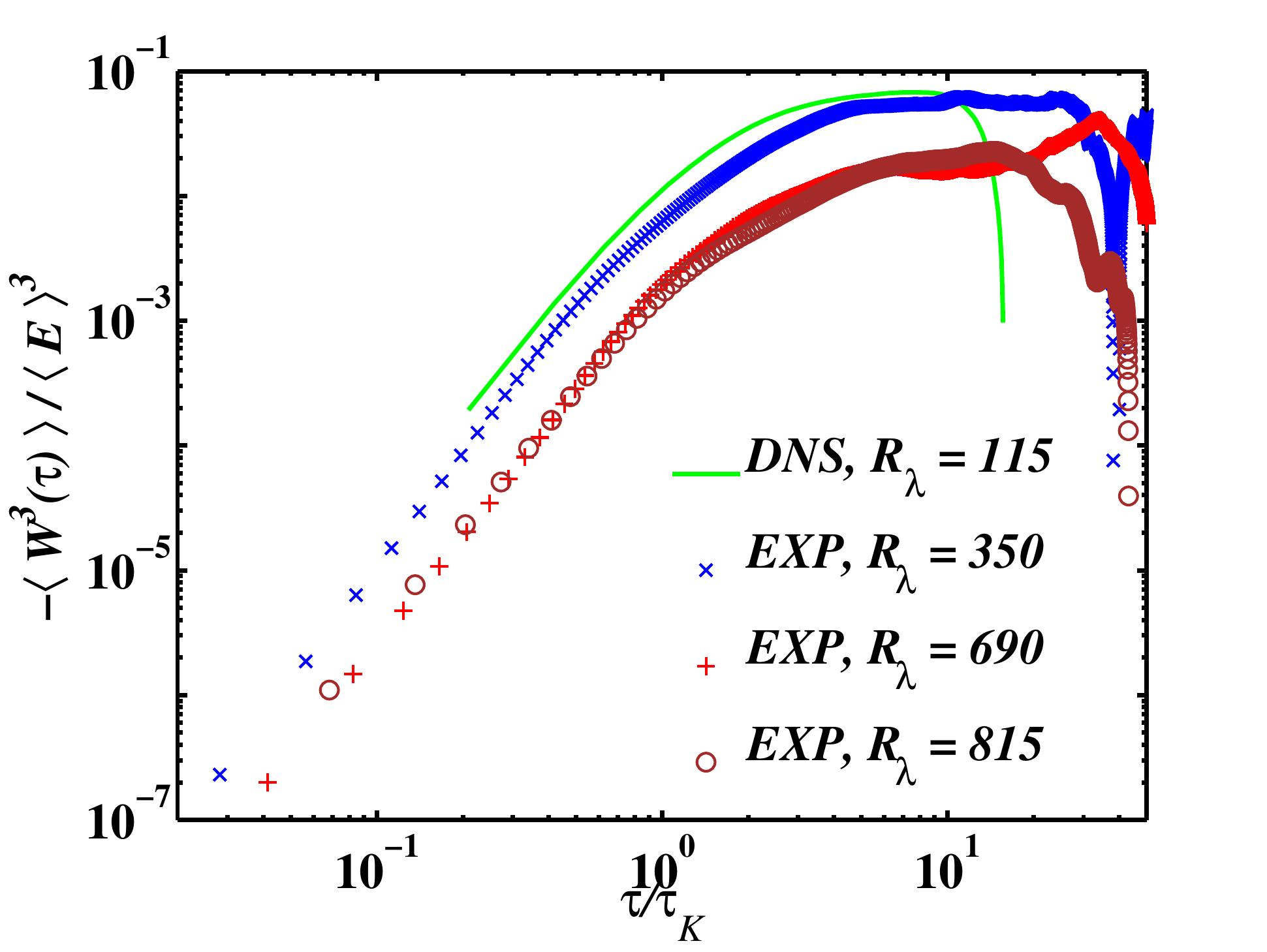}
}
\subfigure[]{
	\includegraphics[width=0.44\textwidth]{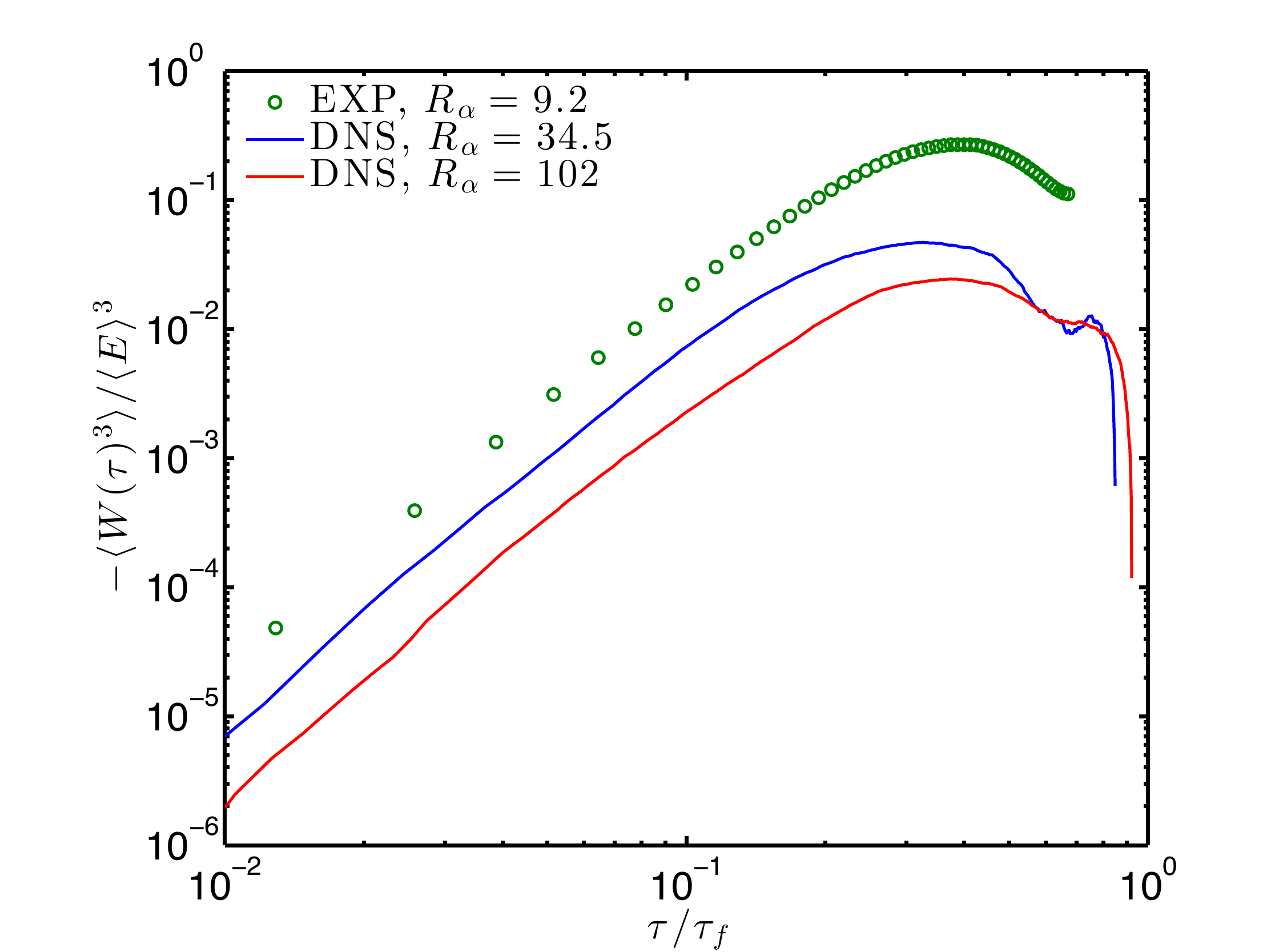}
}
\subfigure[]{
	\includegraphics[width=0.44\textwidth]{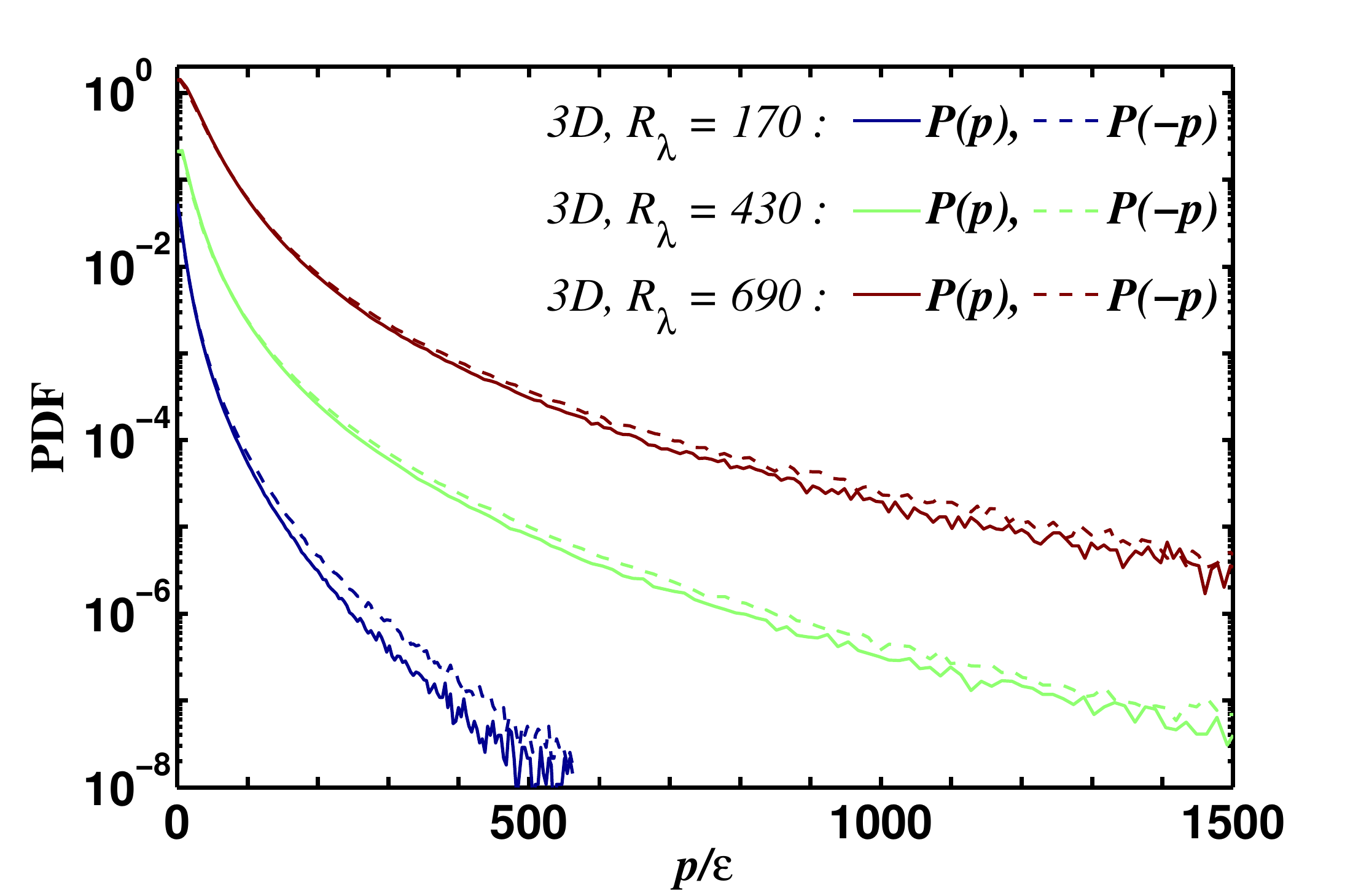}
}
\subfigure[]{
	\includegraphics[width=0.44\textwidth,angle=0]{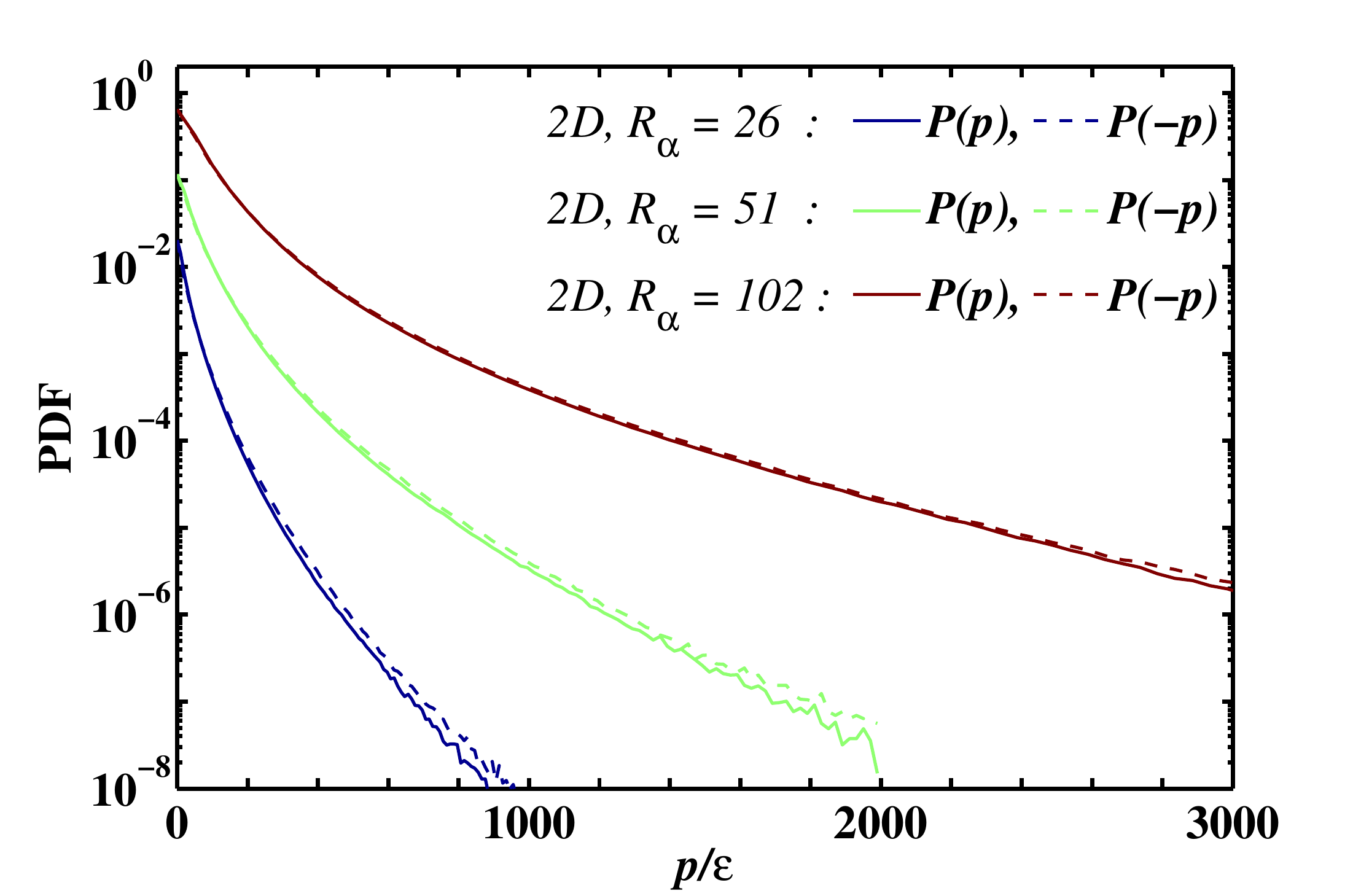}
}
\caption{See the next page for figure caption.}
\label{fig:KE_assym}
\end{center}
\end{figure}

\begin{figure}
\contcaption{
Asymmetry of the statistics of energy differences: (a) The trajectory of a fluid particle in a 3D laboratory flow at
$R_\lambda = 690$. Colour-coding is the instantaneous power $p(t) = \frac{\mathrm{d}E}{\mathrm{d}t} = \mathbf{a}(t)\cdot \mathbf{V}(t)$ acting on the fluid particle, showing that energy builds up slowly and dissipates quickly. The particle enters the observation volume from above and leaves from below.
The indicated length scale is expressed in terms of the Kolmogorov scale $\eta$, which is the dissipation scale of this flow: $l_D = \eta = 30 \mu$m.
(b): The evolution of the kinetic energy $E(t)$ of the same particle as a function of time. The upper panel is for the entire trajectory, while the lower panel magnifies the period with strong energy change, \ie, high power fluctuations (same colour coding as in (a)).
It can be seen that the particle experiences higher values of negative $p$, comparing with positive $p$, indicating that the particle loses kinetic energy more rapidly than gaining kinetic energy.
(c) The third moment of energy increments $W(\tau) = E(t+\tau) - E(t)$ as a function of $\tau$ in 3D turbulence for different Reynolds numbers. The quantity $-\langle W(\tau)^3 \rangle$ grows like
$\tau^3$ at short times, and saturates after $\tau \approx 2 \tau_K$, where $\tau_K$
is the Kolmogorov time, the fastest time scale of turbulence, below which the
viscosity of the fluid damps fluctuations and the dynamics is smooth.
(d) The third moment of $W(\tau)$ from 2D turbulence experiments. Features similar to 3D turbulence are observed, \ie, $W^3(\tau)$ is negative and 
nearly saturates when $\tau \sim \tau_f $, where $\tau_f \sim (l_F^2 / \dissip)^{1/3}$ is the characteristic time corresponding to the forcing scale $l_F$ 
and is much shorter than the time scale over which energy is dissipated by friction (see Appendix).
(e) The PDF of $p/\dissip$ at three different Reynolds numbers $R_\lambda = 170$, $430$, and $690$ for 3D turbulence. For comparison, the PDFs of negative power ($p<0$), shown by the dashed lines, are reflected about the vertical axis.
The size of the characteristic power is much larger than $\dissip$ and increases with the Reynolds number. Moreover, large negative values of $p$ are more frequent than large positive values,
indicating that most of the violent energy exchange events that a fluid
particle experiences are energy-loss events rather than energy-gaining. Data
at $R_\lambda = 170$ and $430$ are from DNS and data at $R_\lambda = 690$ are
from experiments.
(f) PDFs of $p/\dissip$ from 2D turbulence simulations at $R_\alpha = 26$, $51$ and $102$. Similar behaviour as for 3D turbulence is observed.
}
\end{figure}

\clearpage

\begin{figure}
\begin{center}
\subfigure[]{
	\includegraphics[width=0.47\textwidth,angle=0]{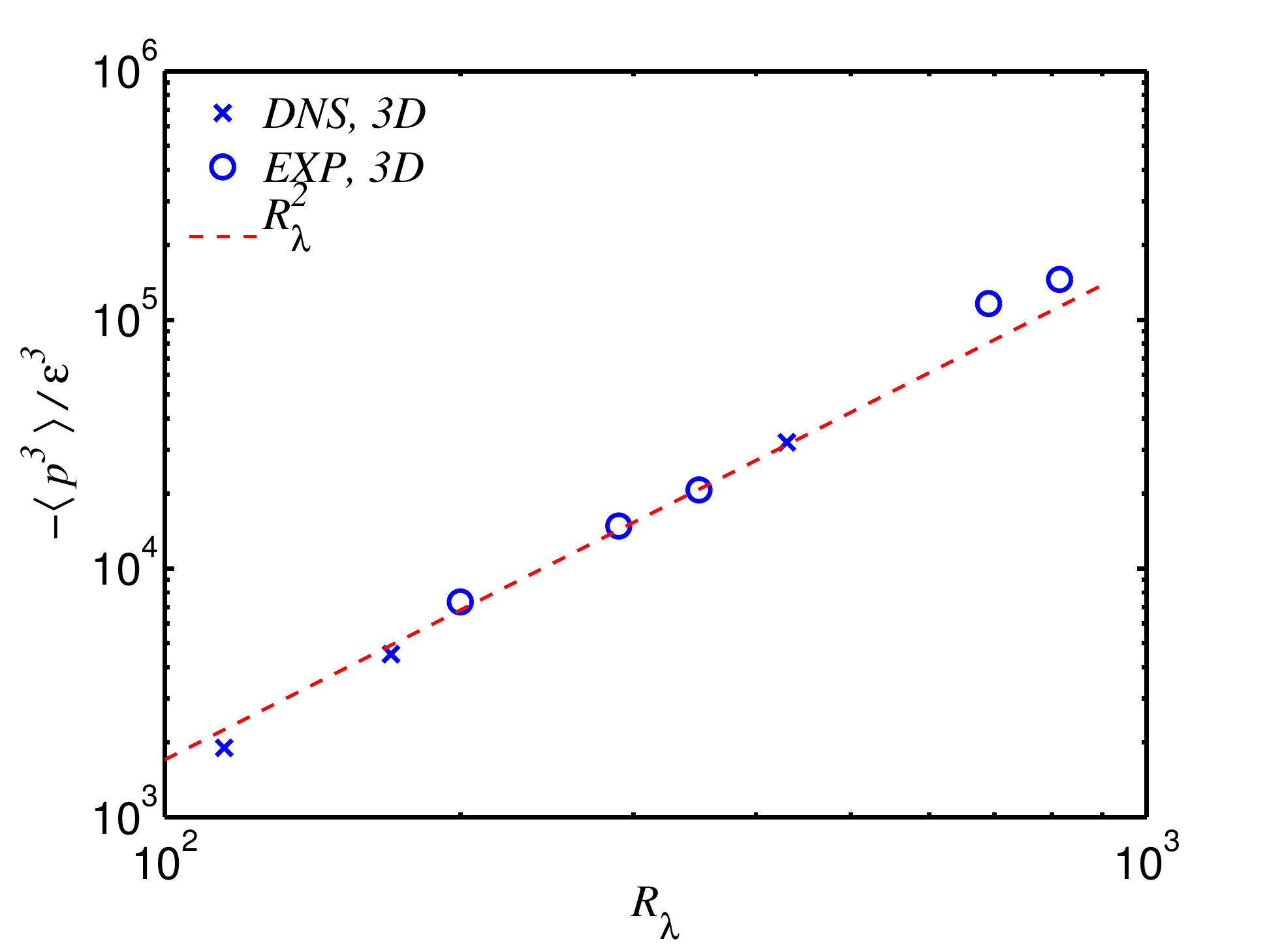}
}
\subfigure[]{
	\includegraphics[width=0.47\textwidth,angle=0]{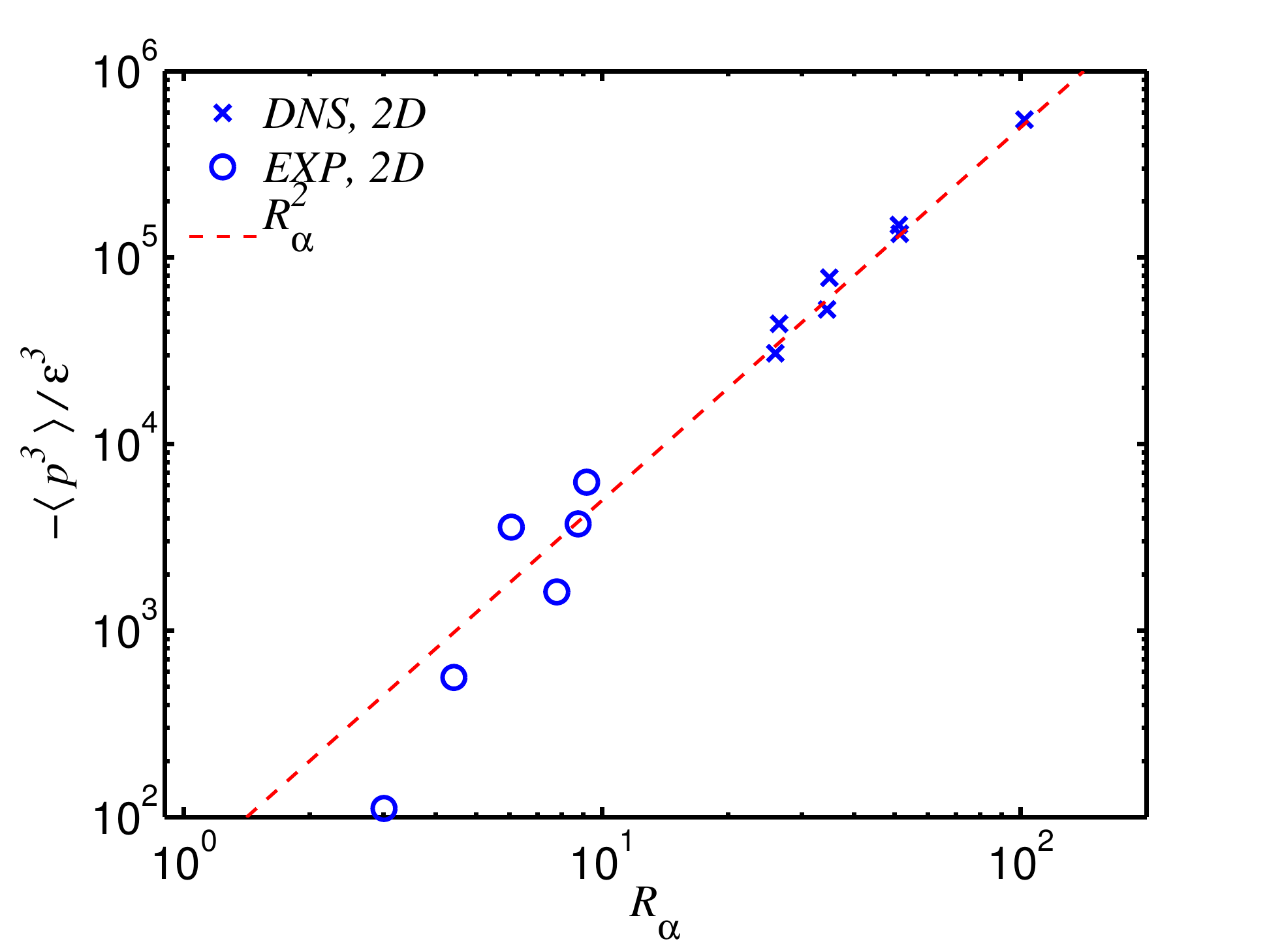}
}
\vspace{5pt}
\subfigure[]{
	\includegraphics[width=0.47\textwidth,angle=0]{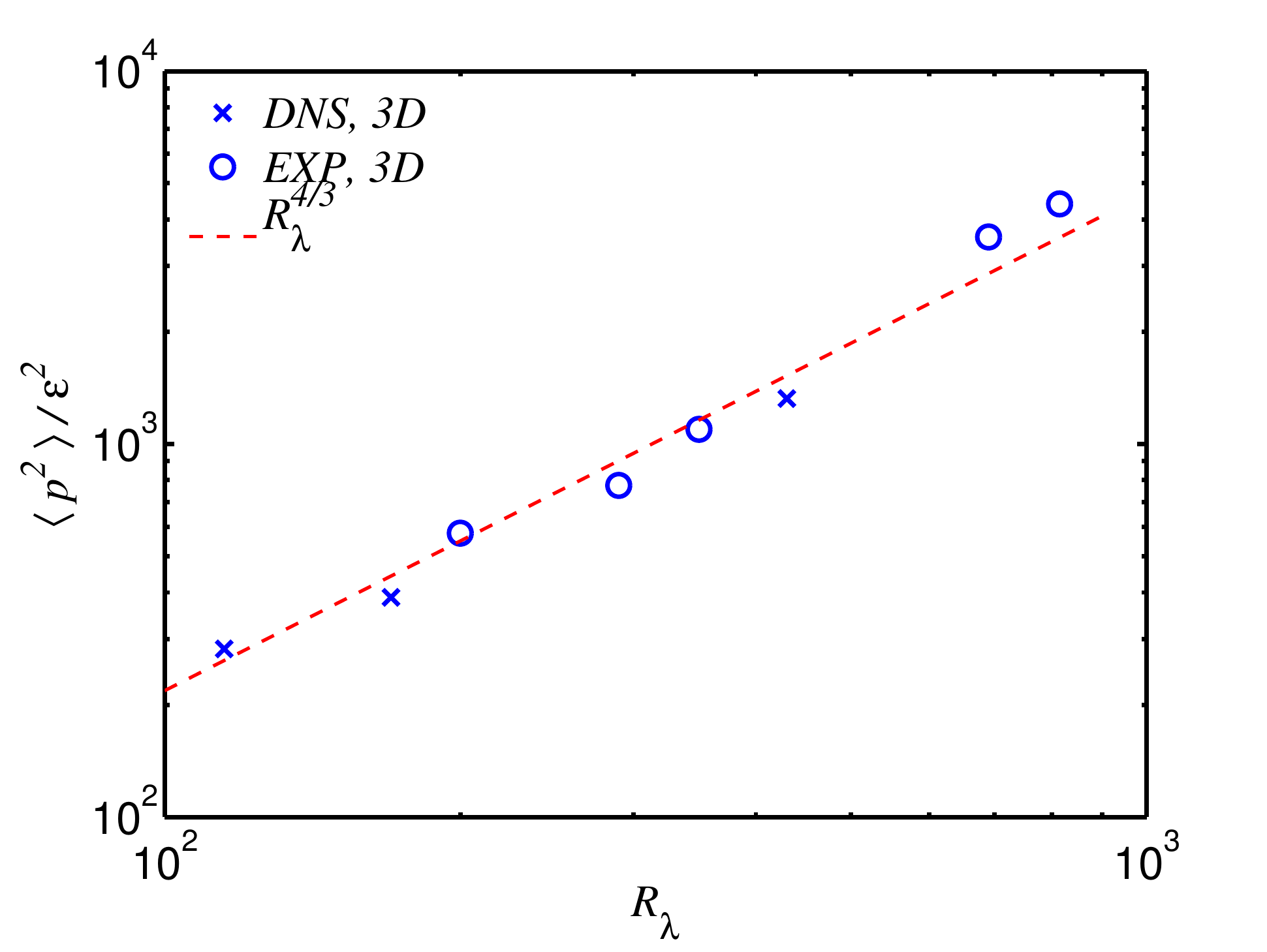}
}
\subfigure[]{
	\includegraphics[width=0.47\textwidth]{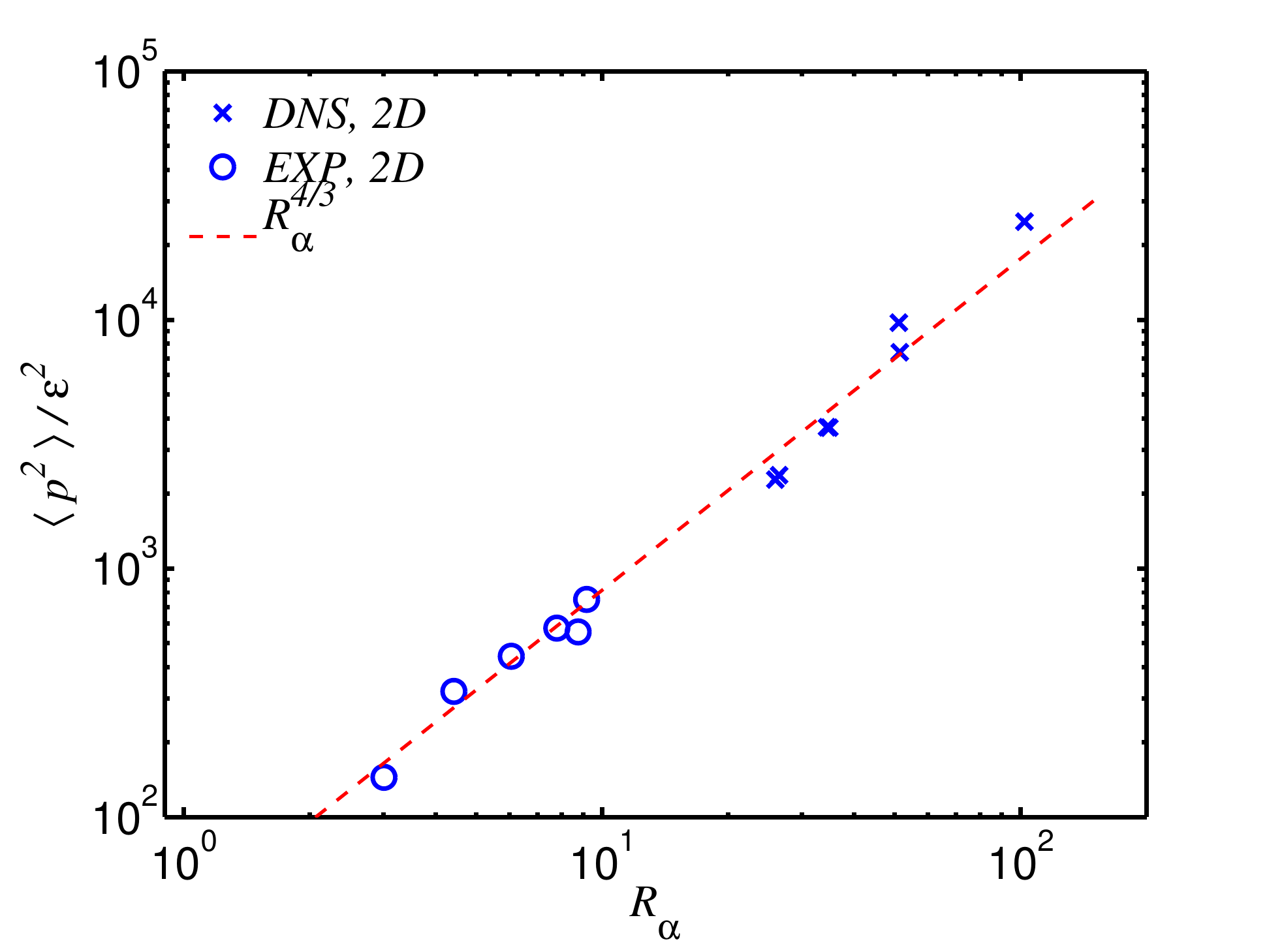}
}
\caption{
Statistical properties of the instantaneous power $p$ acting on fluid particles:
(a): Variation of $ - \langle p^3  \rangle/ \dissip^3$ vs. $R_\lambda$ for 3D turbulence.
Its increase is close to $R_\lambda^2$.
(b): Variation of $ - \langle p^3  \rangle/ \dissip^3$ vs. $R_\alpha$ for 2D turbulence, which increases approximately as $R_\alpha^2$.
(c) and (d): Variation of $\langle p^2 \rangle/\dissip^2$ for 3D and 2D turbulence, respectively. The
variance increases rapidly with Reynolds numbers, close to
$R_\lambda^{4/3}$ or $R_\alpha^{4/3}$ for 3D and 2D turbulence, which results in a nearly constant
skewness for $p$: $\langle p^3\rangle /\langle p^2 \rangle^{ 3/2} \approx -0.5$ in 3D and $\approx - 0.2$ in 2D,
independent of the Reynolds number.
}
\label{fig:W_Stat}
\end{center}
\end{figure}

\clearpage

\begin{figure}
\begin{center}
\subfigure[]{
	\includegraphics[width=0.45\textwidth]{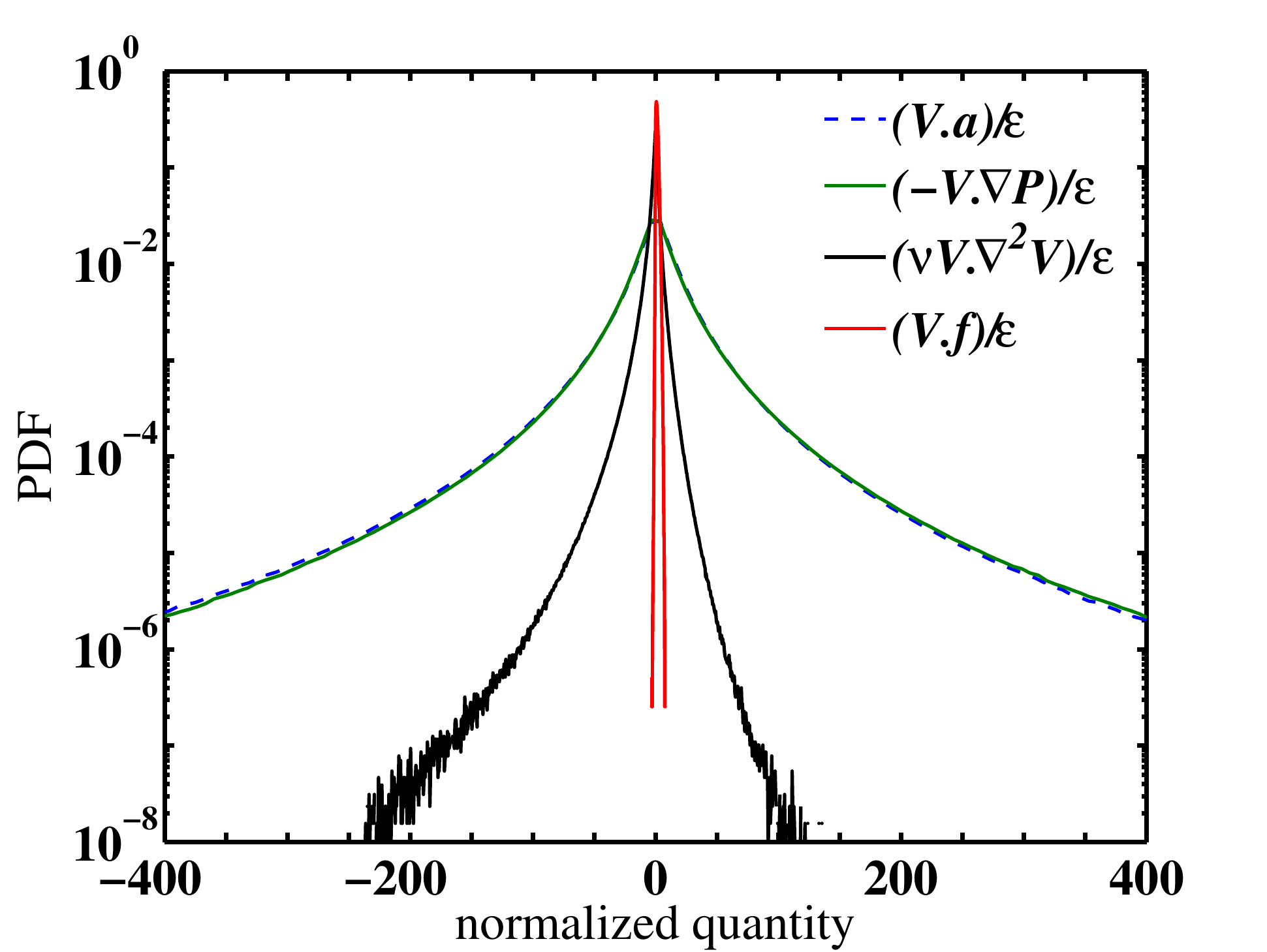}
}
\subfigure[]{
	\includegraphics[width=0.45\textwidth,angle=0]{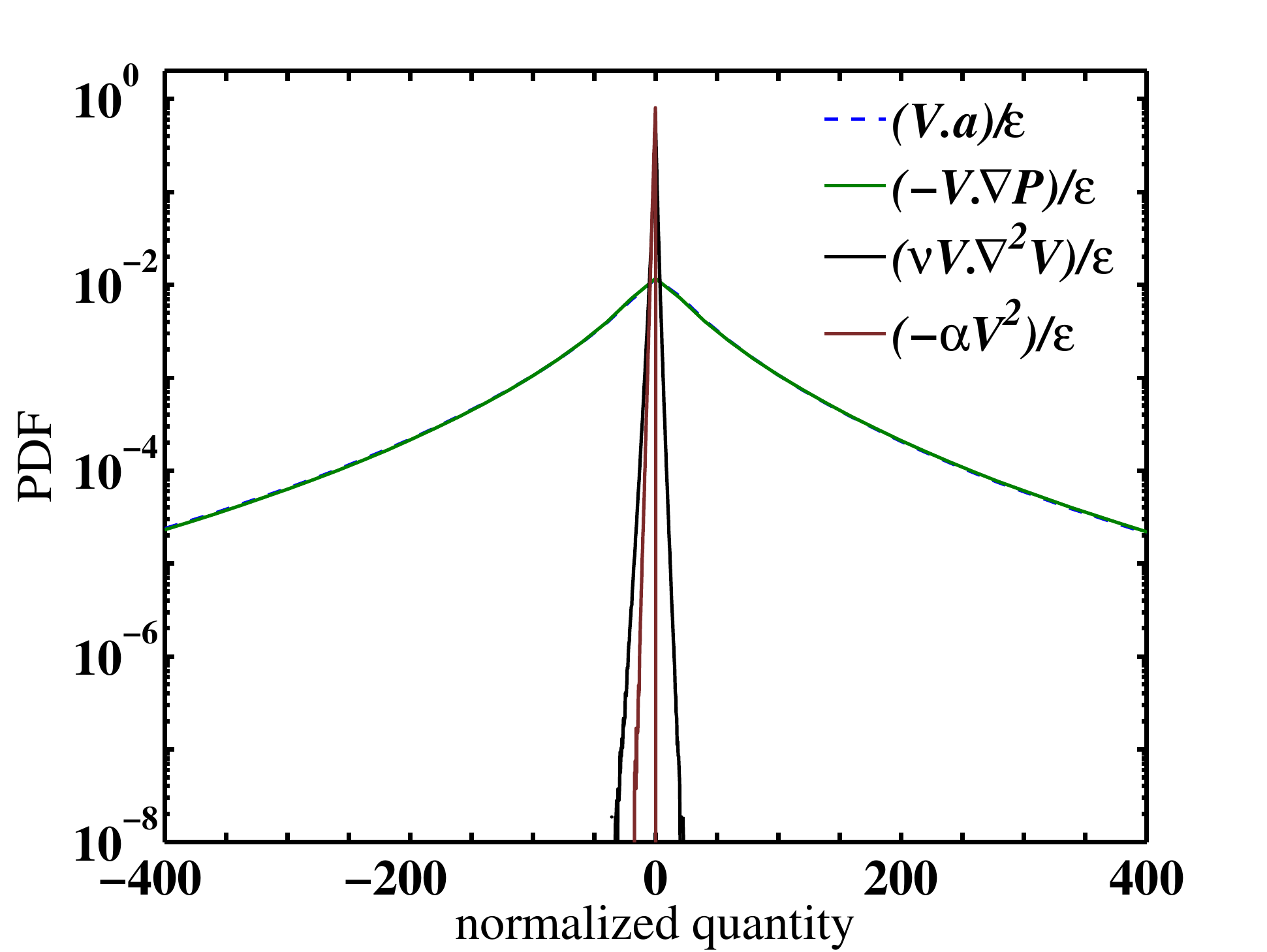}
}
\caption{
Relative importance of the various terms due to forcing, pressure and dissipation on instantaneous power $p = \mathbf{V}\cdot\mathbf{a}$, all normalized with turbulence energy dissipation rate $\dissip$:
(a) The PDFs of $p/ \dissip$,
$ - \mathbf{V} \cdot \nabla P/\dissip$,
$\mathbf{V} \cdot \mathbf{D}/\dissip$ and
$\mathbf{V} \cdot \mathbf{f}/\dissip$
from three-dimensional DNS at $R_\lambda = 430$. The instantaneous values of power and the
contribution from pressure-gradient are typically much larger than those
from the viscous term and the external forcing.
(b) The PDFs of $p/ \dissip$,
pressure gradient term $ - \mathbf{V} \cdot \nabla P/\dissip$,
viscous dissipation term $\mathbf{V} \cdot \mathbf{D}/\dissip$ and
the friction term $-\alpha V^2/\dissip$ from DNS of 2D turbulence at $R_\alpha = 51$.}

\label{fig:NS_contrib}
\end{center}
\end{figure}

\begin{figure}
\begin{center}
\subfigure[]{
	\includegraphics[width=0.48\textwidth,angle=0]{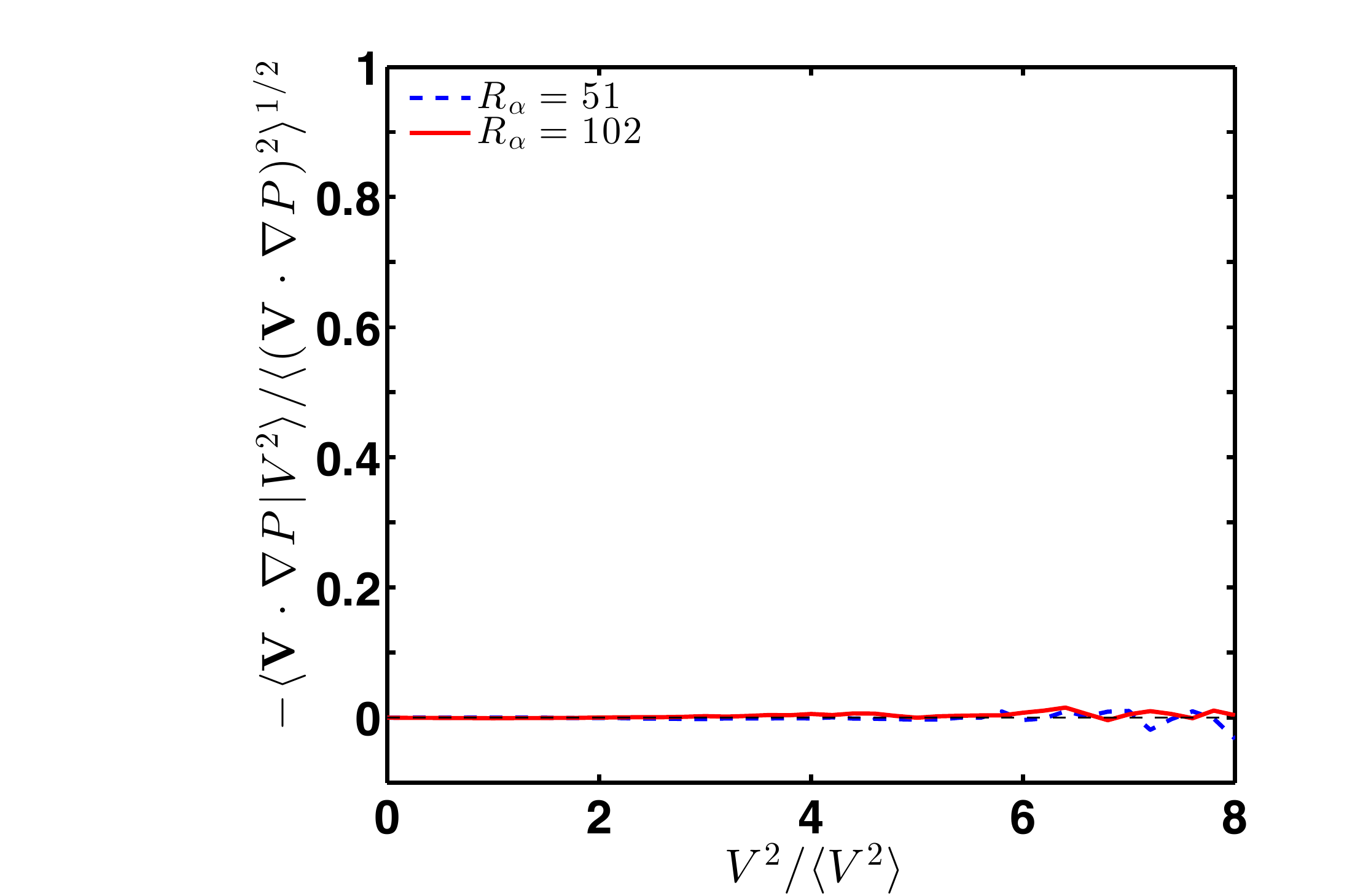}
}
\subfigure[]{
	\includegraphics[width=0.48\textwidth,angle=0]{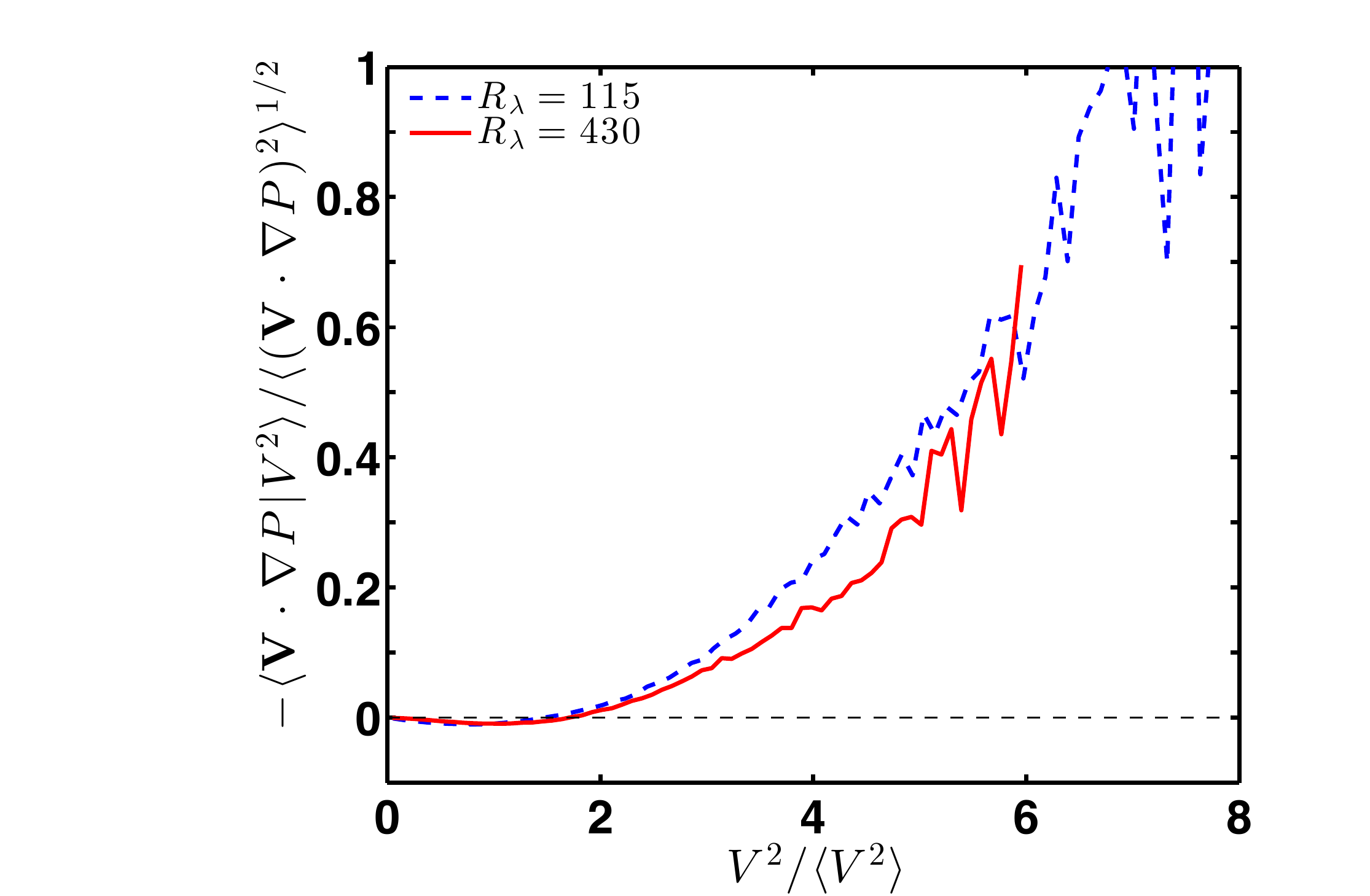}
}
\caption{
 The role of pressure in distributing kinetic energy among fluid particles:
A dramatic difference between 3D and 2D turbulence.
(a) The average the pressure-term $-\mathbf{V} \cdot \nabla P$ conditioned on the kinetic energy
of the fluid particle in 2D turbulence (DNS data at $R_\alpha = 51$ and $R_\alpha = 102$). 
The average $-\mathbf{V} \cdot \nabla P | V^2$ is very small, essentially zero, for all particle velocities,
meaning that in 2D flows, the pressure forces result in no net energy transfer between particles with different levels of kinetic energies,
which is a more ``detailed'' version of the constraint $\langle \mathbf{V}\cdot \nabla P\rangle = 0$.
(b) The same conditional average for 3D turbulence (DNS data at $ R_\lambda =115$ and $430$). Now the situation is very different compared to 2D turbulence. The conditional mean of the pressure is negative for particles with small $V^2$, while it is positive for particles with larger $V^2$,
which implies that, on average, the pressure term in 3D turbulence takes
energy away from {\it slow} particles and gives to {\it fast} particles. This
could lead to a run-away effect and can only be stopped by viscous forces.
}
\label{fig:pressure}
\end{center}
\end{figure}

\begin{figure}
\begin{center}
\includegraphics[width=\textwidth]{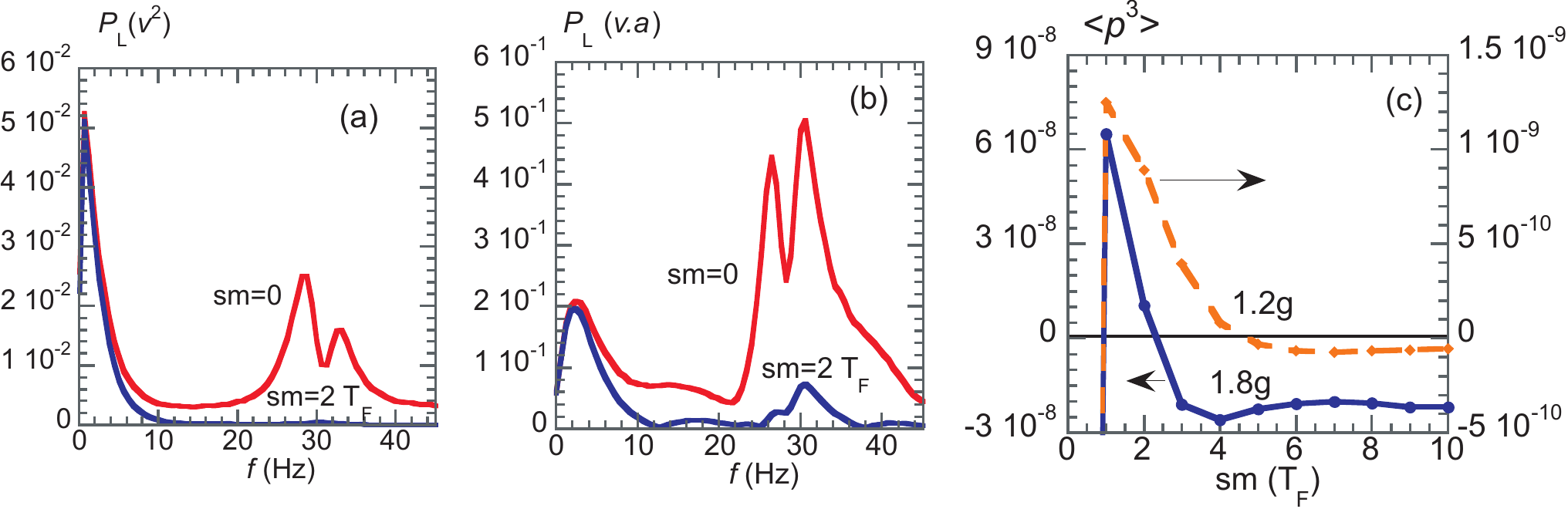}
\end{center}
\caption{Spectral signature of the forcing and the effect of filtering.
Panels (a) and (b) show the Lagrangian spectra of kinetic energy $\mathbf{V}^2$, and the instantaneous power $\mathbf{V} \cdot \mathbf{a}$, respectively.
The curves in red are the raw spectra (without filtering), which exhibit a peak at $f \approx f_0/2$, corresponding to
the forcing, in addition to the peaks at low frequency ($f \approx 2$ Hz) corresponding to the turbulent degrees of freedom.
Low-pass filtering the signals, with a cutoff corresponding to twice the period of the forcing $ sm = 2 T_F$,
considerably reduces the peak at $f_0/2$, as shown by the blue curves, although the peak is  still visible in the spectrum of filtered $\VA$.
(c) Convergence of $\langle p^3 \rangle$, the third moment of the filtered power, as a function of the smoothing parameter $sm$.
}
\label{fig:2d_turb_av_smooth}
\end{figure}

\end{document}